\documentclass[%
reprint,
superscriptaddress,
 amsmath,amssymb,
 aps,
]{revtex4-2}

\usepackage{graphicx}% Include figure files
\graphicspath{{./},{figs/},}

\usepackage{dcolumn}% Align table columns on decimal point
\usepackage{bm}% bold math
\usepackage{physics}
\usepackage{algorithm,algpseudocode,tabularx}

\usepackage[normalem]{ulem}

\usepackage[colorlinks=true,citecolor=blue,linkcolor=magenta]{hyperref}
\usepackage{cleveref}
    \crefname{subsection}{Subsection}{Subsections}

\usepackage[x11names]{xcolor}

\begin{document}

\preprint{APS/123-QED}

\title{Reinforcement learning pulses for transmon qubit entangling gates}

\author{Ho Nam Nguyen}
    % \thanks{Corresponding author: honamnguyen@berkeley.edu}
    \affiliation{Department of Physics, University of California, Berkeley, CA 94720, USA}
    \email{honamnguyen@berkeley.edu}

\author{Felix Motzoi}
\affiliation{Forschungszentrum J\"ulich, Institute of Quantum Control (PGI-8), D-52425 J\"ulich, Germany}

\author{Mekena Metcalf}
\thanks{Current address: HSBC Holdings Plc., One Embarcadero Ctr. San Francisco, CA}
\affiliation{Lawrence Berkeley National Laboratory, Berkeley, CA 94720, USA}

\author{K.~Birgitta Whaley}
\affiliation{Department of Chemistry, University of California, Berkeley, CA 94720, USA}

\author{Marin Bukov}
    \thanks{These authors contributed equally}
    \affiliation{Max Planck Institute for the Physics of Complex Systems, N\"othnitzer Str. 38, 01187 Dresden, Germany}
    \email{mgbukov@pks.mpg.de}
\author{Markus Schmitt}
    \thanks{These authors contributed equally}
    \affiliation{Forschungszentrum J\"ulich, Institute of Quantum Control (PGI-8), D-52425 J\"ulich, Germany}
    \affiliation{University of Regensburg, 93053 Regensburg, Germany}

\date{\today}% It is always \today, today,

\begin{abstract}
The utility of a quantum computer is highly dependent on the ability to reliably perform accurate quantum logic operations. For finding optimal control solutions, it is of particular interest to explore model-free approaches, since their quality is not constrained by the limited accuracy of theoretical models for the quantum processor -- in contrast to many established gate implementation strategies.
In this work, we utilize a continuous control reinforcement learning algorithm to design entangling two-qubit gates for superconducting qubits; specifically, our agent constructs cross-resonance and CNOT gates without any prior information about the physical system. 
Using a simulated environment of fixed-frequency fixed-coupling transmon qubits, we demonstrate the capability to generate novel pulse sequences that outperform the standard cross-resonance gates in both fidelity and gate duration, while maintaining a comparable susceptibility to stochastic unitary noise. We further showcase an augmentation in training and input information that allows our agent to adapt its pulse design abilities to drifting hardware characteristics, importantly, with little to no additional optimization.
Our results exhibit clearly the advantages of unbiased adaptive-feedback learning-based optimization methods for transmon gate design.
\end{abstract}

\maketitle
%\tableofcontents

\section{\label{sec:intro}Introduction}

% \protect
% \red[]{make an arxiv version wihout texts even in tex file} 
Quantum computing holds immense potential to revolutionize various fields, such as optimization, simulation, and cryptography – in some cases promising exponential computational speedup compared to its classical counterpart. However, a central obstacle to harnessing this potential is the challenge of realizing reliable quantum operations. Therefore, achieving high-fidelity quantum gates is a crucial prerequisite to unlocking the full potential of quantum computing for practical applications.

A common approach is to optimize control protocols based on effective models of the physical platform. With a suitable model at hand, high-fidelity strategies can often be achieved through analytical insights,  gradient-based optimization methods, or error amplification techniques~\cite{freeman1998spin,guery2019shortcuts,khaneja2005optimal,de2011second, sorensen2018quantum,erroramp_lucero_2010}. However, present-day quantum systems are characterized by substantial levels of noise, decoherence, and other environmental disturbances, for which accurate models are rarely known. Moreover, even when a good model is known, it is often not exactly solvable, limiting its usefulness for optimal quantum gate design. A route to circumvent these shortcomings is to resort to model-free approaches, which facilitate gate optimization through direct interactions with the quantum device.

Recently an adaptive approach using reinforcement learning (RL) has become increasingly popular in quantum gate design due to its model-free nature, obviating the need for a precise description of all details of the system~\cite{aiquantum_kren_2023}. When trained in a simulation, the RL approach rivals optimal control techniques in synthesizing high-precision quantum gates using discretized control for qubit-based~\cite{An_2019_drlbangbang, shindi2023modelfree_bangbang,Dalgaard_2020_alphazero} and qudit-based~\cite{An_2021_drlbangbang_multilevel_dissipative,rlgkp_sivak_2022} systems, including potentially drastic improvements in exploration \cite{Dalgaard_2020_alphazero} and sample efficiency~\cite{sampleefficiency_khalid_2023}.
Similarly, RL was found successful in optimizing continuous controls for a generic qubit model~\cite{HuChenDong2022TD3Gatedesign} as well as a hardware-specific gmon model~\cite{niu2018ufo}. Ref.~\cite{niu2018ufo} further demonstrates the resilience of the RL-designed pulse sequences to stochastic noise when optimized with knowledge of a noisy environment. Discrete control algorithms have also been adapted to successfully learn faster single-qubit gates from scratch using experimental data from IBM's superconducting platform~\cite{wright2023fast,Baum2021QCtrl};  Ref.~\cite{Baum2021QCtrl} additionally uses RL to improve upon the standard structure of an analytical cross-resonance pulse sequence.

While many aspects of the RL approach are more broadly applicable, in this work, we will specifically address the gate design problem for fixed-frequency fixed-coupling transmon qubits. For this platform, the model-based approach has yielded valuable insights in the pursuit of crafting high-fidelity entangling gates by utilizing the cross-resonance interaction~\cite{crgate_paraoanu_2006,crgate_rigetti_2010}. An effective approximate analytical model, capable of capturing the dominant Hamiltonian terms generated by the primary cross-resonance drive as well as undesired cross-talk~\cite{Magesan_2020}, has paved the way for the development of various error suppression techniques, including echo sequences~\cite{Echo_Corcoles_2013}, selective-darkening/active cancellation~\cite{sdgate_deGroot_2010,sdcnot_deGroot_2012,Sheldon_2016}, optimal control theory \cite{kirchhoff2018optimized}, rotary pulsing~\cite{Rotary_Sundaresan_2020}, and most recently derivative pulse shaping \cite{li2023suppression}. However, the intricate nature of real hardware and its inevitable imperfections persist, hindering our ability to achieve flawless quantum gate operations. Moreover, while the analytical insight motivates a specific family of control pulses, it remains unknown whether even better solutions can be found by expanding the considered protocol space.

As mentioned above, RL approaches have been explored to design entangling gates for the transmon platform. However, one of the inherent strengths of RL -- its capability to discover innovative strategies free from the confines of theoretical protocol sequences -- has remained underutilized. The absence of such flexibility results in lengthy pulse sequences which, in turn, impose severe limitations on the fidelity of these operations. Moreover, the ability of RL agents to learn adaptive strategies, that include optimal reactions to the feedback received when they are deployed, has so far received little attention. For example, although RL solutions display a degree of temporal robustness due to exposure to changing underlying system characteristics during training~\cite{Baum2021QCtrl}, leveraging the adaptability of the RL agent to deal with such fluctuations explicitly remains largely unexplored.

In this work, we address these and related open questions by deploying a continuous control RL algorithm to construct piece-wise constant (PWC) pulse sequences for cross-resonance and CNOT gates without any prior knowledge about the controlled system. We emphasize that this model-free approach only requires feedback from the environment (simulated or experimental) and has no information about the physical model for the environment’s dynamics. Our RL training agent only has access to the quantum state and the gate fidelity, which, in principle, can be obtained experimentally via tomography and fidelity benchmark; however, in this work we train the RL agent in simulation. We tailor the simulated environment to fixed-frequency fixed-coupling transmons using realistic system characteristics to have a direct comparison between our RL results and the existing error suppression techniques in superconducting platforms. Note that in this particular transmon architecture, the qubit frequency depends on the fabrication of the transmon chip itself and cannot be controlled throughout the gate duration.

We first demonstrate that our unbiased RL agent is capable of generating novel high-fidelity control solutions that outperform current state-of-the-art cross-resonance pulse sequences. By effectively navigating the vast design space of multi-segment PWC functions to identify high-quality pulses for multiple continuous control drives simultaneously, our agent has achieved a remarkable feat in addressing an up to 120-dimensional control problem, as compared to the 20-dimensional problem considered in Ref.~\cite{Baum2021QCtrl}. Without compromising the fidelity, our agent additionally discovers control solutions with large drive amplitudes, that can lead to a maximum reduction in gate duration by 30\%, while being feasible to implement on modern NISQ devices. We further show how to augment our RL approach so that our agent can learn to adapt to drifts in the underlying hardware parameters (characteristics), a common issue that plagues near-term superconducting devices. This adaptation offers a two-fold advantage: immediate, high-fidelity control solutions without any extra optimization when dealing with moderate drifts or a reduction in the number of training iterations required to address more significant changes in hardware parameters. These findings underscore the practicality of the RL approach as a potent alternative for tackling the quantum gate design problem.

The following sections are organized as follows. We start with defining the quantum gate design problem for one and two-qubit gates in Sec.~\ref{sec:design_problem}. We give a brief overview of the state-of-the-art gate implementations in Sec.~\ref{sec:default_implementation}.  We then present our reinforcement learning approach in Sec.~\ref{sec:rl} and our simulated results in  Sec.~\ref{sec:results}. Finally in Sec.~\ref{sec:discussion}, we conclude and discuss future directions.

\section{Quantum gate design problem}\label{sec:design_problem}

A quantum gate design task aims to realize a logical operation over one or more qubits via optimizing a set of available time-dependent external control fields $\{d_j(t)\}$ over some gate duration $T$. The effect of these fields is described by a control Hamiltonian $H_{\rm ctrl}(t)=H_{\rm ctrl}[\{d_j(t)\}]$ while the intrinsic dynamics of the qubits is captured by the system Hamiltonian $H_{\rm sys}$. Together, they generate the full unitary evolution
\begin{equation}
    U = \mathcal{T}\exp\left[-i\int_0^T\left(H_{\rm ctrl}(t)+H_{\rm sys}\right)\mathrm dt\right],
\end{equation}
where $\mathcal{T}$ denotes the time-ordering operator. We measure the accuracy of approximation of the target operation $U_{\rm target}$ by the resultant unitary $U$ via the average gate fidelity \cite{Pedersen_2007} 
\begin{eqnarray}
    \mathcal{F}_{\rm avg}(U,U_{\rm target})
    &=& \int \mathrm d\psi_0 \left|\ev{U^{\rm qubit}U_{\rm target}^{\dagger}}{\psi_0}\right|^{2}  \nonumber \\
    &=& \int \mathrm d\psi_0  \left|\ev{M}{\psi_0}\right|^{2} \nonumber \\
    &=& \frac{\Tr(MM^{\dagger}){+}|\Tr (M)|^{2}}{n(n+1)}, \label{eq:Favg}
\end{eqnarray}
where $U^{\rm qubit}=\Pi^{\rm qubit} U \Pi^{\rm qubit}$ is the unitary map projected to the qubit subspace of dimension $n$. The average is taken over all initial states $\ket{\psi_0}$ distributed uniformly according to the Haar measure. Here we focus on superconducting qubit platforms where local $Z$ rotations can be performed virtually~\cite{virtualZ_McKay_2017}, i.e., without incurring any additional time. We include this degree of freedom in the unitary by augmenting $U^{\rm qubit}$ to $V_Z(\bm{\theta}) U^{\rm qubit}$, in which the near-optimal angles $\bm{\theta}$ are given by the matrix elements of $M=U^{\rm qubit}U^\dagger_{\rm target}$, see App.~\ref{app:virtualZ}. 

We consider a target gate fidelity of 99.9\%, which is an order of magnitude higher than the 99\% fidelity of the surface code threshold for two reasons. First, this level of fidelity is expected to enjoy a drastic reduction in the number of physical qubits when using the surface code~\cite{surfacecode}. Second, for the typical two-qubit gate durations considered in this work, e.g., 248.9 ns and 177.8 ns (the smallest time unit is the inverse sampling rate $dt=2/9$ ns for the considered device), the gate fidelity is coherence limited at 99.9\% and 99.93\%, respectively. These limits are determined by computing the average gate error under a channel with the amplitude damping rate $T_1=300\mu$ s and the phase damping rate $T_2=300\mu$s~\cite{coherencelimit_kenxuan2024}, which are achievable in current devices~\footnote{https://quantum-computing.ibm.com/}. Thus, having in mind any realistic decoherence in the near term, our target fidelity of 99.9\% coincides with the coherence limited fidelities for the considered range of gate duration.

In addition to the average gate fidelity, we also investigated the worst-case fidelity as an alternative figure of merit. However, we did not find any discernible advantage and report this additional result in App.~\ref{app:worstcase}. 

In the following, we provide the explicit form of the Hamiltonian used to model superconducting transmon qubits, while state-of-the-art gate implementations are discussed next in Sec.~\ref{sec:default_implementation}.

\subsection{Single-qubit Hamiltonian}\label{subsec:singlequbitgates}

We begin by modeling a single transmon in the Duffing approximation \cite{khani2009duffing}, with the lab frame Hamiltonian ($\hbar=1$): 
\begin{equation}
    H_{\rm 1,sys}^{\rm lab} = \omega b^{\dagger}b+\frac{\alpha}{2}b^{\dagger}b^{\dagger}bb, \label{eq:H1sys}
\end{equation}
where $\omega$ and $\alpha$ denote the $\ket{0}\leftrightarrow\ket{1}$ transition frequency and anharmonicity, respectively, and $b,b^\dagger$ are ladder operators. This transmon can be driven at frequency $\omega_d$ via a control Hamiltonian
\begin{eqnarray}
    H_{\rm 1,ctrl}^{\rm lab}(t) &=& \Omega_d\Re\left(d(t)e^{i\omega_d t}\right)(b^{\dagger} + b), \label{eq:H1ctrl}
\end{eqnarray}
where we have factored out the drive strength $\Omega_d$ to keep the real and imaginary parts of the complex control signal $d(t)$ normalized to $[-1,1]$. By rotating to the driving frame via the transformation $R(t)=e^{-i\omega_d t b^{\dagger}b}$ and ignoring fast rotating terms (cf.~App.~\ref{app:transformation} for details), we arrive at the rotating frame Hamiltonian.
\begin{eqnarray}
    H_1(t) &=& R^{\dagger}(t)(H_{\rm 1,sys}^{\rm lab} + H_{\rm 1,ctrl}^{\rm lab}(t)-i\partial_t)R(t) \nonumber \\
    &\approx&  \delta b^{\dagger}b+\frac{\alpha}{2}b^{\dagger}b^{\dagger}bb + \left[\frac{\Omega_d}{2}d(t)b + {\rm h.c.}\right], \label{eq:H1}
\end{eqnarray}
where $\delta=\omega-\omega_d$ is the qubit detuning, which vanishes when the transmon is driven on resonance, i.e., $\omega_d = \omega$.  

To make the connection to single-qubit rotations explicit, we consider only the first two levels and make the replacement $b\rightarrow \sigma^-=X-iY$ and $b^{\dagger} \rightarrow \sigma^+=X+iY$ in the control Hamiltonian, which results in
\begin{equation}
    H_{\rm 1,ctrl}^{\rm qubit}(t) = \frac{\Omega_d}{2}\Re(d(t)])X + \frac{\Omega_d}{2}\Im(d(t))Y, \label{eq:Hctrl_1}
\end{equation}
where $X$ and $Y$ denote respectively the Pauli-$X$ and Pauli-$Y$ matrices. Evidently, turning on the complex control field $d(t)$ induces qubit rotation around the $x$ and $y$ axes, which, for long enough gate duration, is sufficient for realizing any single-qubit gate. With the Euler-angle decomposition,
\begin{equation}
    U_3(\theta,\phi,\lambda) = V_Z(\phi) R_X\left(-\frac{\pi}{2}\right) V_Z(\theta) R_X\left(\frac{\pi}{2}\right) V_Z(\lambda),
\end{equation}
one can achieve any desired single-qubit gate by merely calibrating the $R_X(\pm \pi/2)$ rotations, and, as discussed above, $Z$ gates can be implemented virtually. Note that in practice much shorter gate durations are desirable, in which case a number of errors (such as high state population \cite{Wood_2018_leakage}) will unavoidably arise and therefore need to be counteracted, see Sec.~\ref{sec:default_implementation}.

\subsection{Two-qubit Hamiltonian}\label{subsec:2transmon}

\begin{figure}[t!]
    \centering
    \includegraphics[width=0.48\textwidth]{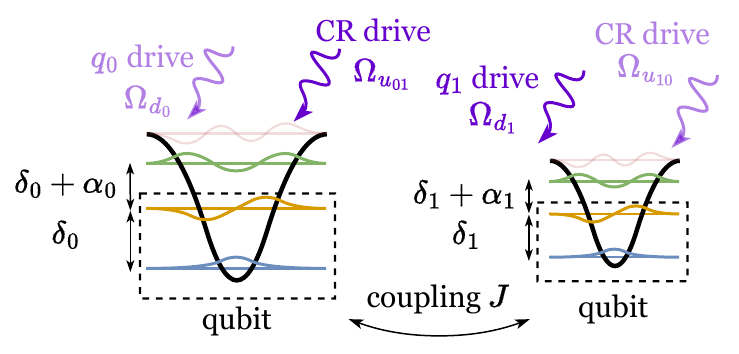}
    \caption{\textbf{Coupled transmons simulated as Duffing oscillators.} Simulation is truncated at three energy levels per transmon (the faded fourth level is shown but not considered), and performed in a rotating frame. The first two levels act as qubits (dashed boxes). External control drives (purple) include on-resonance and cross-resonance complex control fields, denoted by $d(t)$ and $u(t)$ in the main text, respectively. The full Hamiltonian in~Eq.~\ref{eq:H2total} is completely characterized by the detuning $\delta_j$ and anharmonicity $\alpha_j$ for each transmon, drive strengths $\{\Omega_{d_0},\Omega_{u_{01}},\Omega_{d_1},\Omega_{u_{10}}\}$ for 4 external controls, and the direct coupling $J$.
    }
    \label{fig:transmons}
\end{figure}

We now extend the model to describe a pair of transmons by combining two Duffing Hamiltonians coupled by a resonator. The resonator acts as a bus for coherent communication between the quantum states of the two Hamiltonians that will define the qubits. Although a variety of different logical two-qubit gates are possible with this setup \cite{goerz2017charting}, we will discuss here the cross-resonance interaction Hamiltonian, which is the current standard for fixed-frequency architecture.

When the resonator's fundamental frequency is much larger than both $\ket{0}{\leftrightarrow}\ket{1}$ transition frequencies of the individual transmons, we can project the Hamiltonian onto the zero-excitation subspace of the bus resonator to obtain the following lab-frame effective Hamiltonian:
\begin{equation}
    H_{{\rm 2,sys}}^{\rm lab}=\sum_{j=0}^{1}\left(\Tilde{\omega}_{j}b_{j}^{\dagger}b_{j}+\frac{\alpha_{j}}{2}b_{j}^{\dagger}b_{j}^{\dagger}b_{j}b_{j}\right)+J\left(b_{0}^{\dagger}b_{1}+b_{0}b_{1}^{\dagger}\right).
\end{equation}
Here, $\Tilde{\omega}_j$ denotes the resonator-dressed qubit frequency and $J$ denotes the effective coupling strength~\cite{Magesan_2020}.

In addition to a standard on-resonance control field $d(t)$ on each transmon, an entangling operation can be realized by driving one qubit at the frequency of another via a cross-resonance (CR) control field $u(t)$. The two-transmon control Hamiltonian then becomes
\begin{eqnarray}
    H_{\rm 2,ctrl}^{\rm lab}(t) &=& \Omega_{d_0} \Re(e^{i\Tilde{\omega}_{0}t}d_{0}(t))(b^\dagger_0+b_0) \nonumber \\
    &&+ \Omega_{u_{01}} \Re(e^{i\Tilde{\omega}_{1}t}u_{01}(t))(b^\dagger_0+b_0) \nonumber \\
    &&+ \Omega_{d_{1}} \Re(e^{i\Tilde{\omega}_{1}t}d_{1}(t))(b^\dagger_1+b_1) \nonumber \\
    &&+ \Omega_{u_{10}} \Re(e^{i\Tilde{\omega}_{0}t}u_{10}(t))(b^\dagger_1+b_1).
\end{eqnarray}
Here, $u_{01}(t)$ refers to the cross-resonance pulse sent to qubit $0$ when driven at the frequency of qubit $1$, and vice versa for $u_{10}(t)$.

Moving into the frame rotating at $\omega_d$ for both transmons using the transformation $R(t)=e^{-i\omega_d t(b_0^{\dagger}b_0+b_1^{\dagger}b_1)}$, and ignoring the fast rotating terms, we obtain the rotating-frame Hamiltonian 
\begin{eqnarray}
    H_2(t) &=& R^{\dagger}(t)(H_{\rm 2,sys}^{\rm lab} + H_{\rm 2,ctrl}^{\rm lab}(t)-i\partial_t)R(t) \nonumber \\
    &=&  \sum_{j=0,1}\left(\delta_j b_j^{\dagger}b_j+\frac{\alpha_j}{2}b_j^{\dagger}b_j^{\dagger}b_jb_j\right) + J\left(b_{0}^{\dagger}b_{1}+b_{0}b_{1}^{\dagger}\right) \nonumber \\
    &+& \frac{b_0}{2}\left[\Omega_{d_0}e^{i\delta_{0}t}d_{0}(t)+\Omega_{u_{01}}e^{i\delta_{1}t}u_{01}(t)\right]+{\rm h.c.} \nonumber \\ 
    &+&\frac{b_1}{2}\left[\Omega_{d_1}e^{i\delta_{1}t}d_{1}(t)+\Omega_{u_{10}}e^{i\delta_{0}t}u_{10}(t)\right]+{\rm h.c.} \label{eq:H2total}
\end{eqnarray}
where $\delta_j=\Tilde{\omega}_j-\omega_d$ defines the detuning for the $j$-th transmon. The pair of transmons, whose dynamics is described by $H_2(t)$, is illustrated in Fig.~\ref{fig:transmons}. In this work, we simulate the dynamics in the frame rotating at the second transmon's frequency, i.e., setting $\delta_1=0$. 

With the first transmon as control and the second as target, the main effect of the cross-resonance drive can be studied by setting $u_{01}$ to a constant value $\Omega$ and other control fields to zero:
\begin{eqnarray}
    H_{{\rm 2}}^{\rm CR}(t) &=& \delta_0 b_0^{\dagger}b_0+ \sum_{j=0,1}\frac{\alpha_j}{2}b_j^{\dagger}b_j^{\dagger}b_jb_j \nonumber \\ 
    &+& J\left(b_{0}^{\dagger}b_{1}+b_{0}b_{1}^{\dagger}\right) + \frac{\Omega}{2}u_{01}(t)b_0 + \text{h.c}. \label{eq:H2CR}
\end{eqnarray}

To obtain the effective $ZX$ interaction rate (or strength) within the qubit subspace while accounting for higher levels, one can employ perturbation theory~\cite{Magesan_2020} to obtain the following approximate effective CR Hamiltonian
\begin{equation}
    H^{\rm CR}_{\rm eff} = \sum_{A,B} \omega_{AB} A\otimes B, \label{eq:CRHamiltonian}
\end{equation}
where $A \in \{I,Z\}  $ and $B\in \{I,X,Z\}$. In the presence of classical cross-talk and incorrect phases in control drives, $B$ can be extended to include the Pauli$Y$ matrix~\cite{Sheldon_2016}. Within perturbation theory for small coupling $J$ and small drive $\Omega$, the interaction rates have the following scaling
\begin{eqnarray}
    \omega_{ZX},\omega_{IX}&\sim& \Omega, \qquad
    \omega_{ZI},\omega_{ZZ}\sim \Omega^2. 
    \label{eq:interactionrate}
\end{eqnarray}
The resultant dominant $ZX$ term can then be used to implement the following entangling operation
\begin{equation}
    ZX(\pi/2) = \exp\left(-i\frac{\pi}{4}ZX\right)= \frac{I-iZX}{\sqrt{2}},
\end{equation}
known as the cross-resonant (CR) gate, which is locally equivalent to the popular CNOT gate. Such entangling operations, together with the capacity to realize any single-qubit gate, enable universal quantum computation in the superconducting transmon platform.

\subsection{Leakage}

Although only the first two levels of a transmon are used to represent a qubit, the higher levels are nevertheless still present and can be populated as the system evolves. We capture the most prominent leakage outside the ideal computational qubit subspace by including the second excited state of the anharmonic oscillator $\ket{2}$ (cf.~Fig.~\ref{fig:transmons}). The full state space can be decomposed into a direct sum of the computational subspace $\chi_1$ and the leakage subspace $\chi_2$. Projectors onto these subspaces respectively are denoted as $I_1$ and $I_2$~\footnote{The two-qutrit states are ordered as $(00,01,02,10,...)$, the projectors are given by $I_1=\text{diag}(1,1,0,1,1,0,0,0,0)$ and $I_2=\text{diag}(0,0,1,0,0,1,1,1,1)$}. 

Under a unitary quantum channel $\mathcal{E}(\rho)=U\rho U^\dagger$, state leakage averaged over all initial pure states in the qubit subspace is given by
\begin{equation*}
    L = \int \mathrm d\psi_0 \Tr \left[I_2 \mathcal{E}(\dyad{\psi_0}) \right] 
    = \Tr\left[I_2\mathcal{E}\left(\frac{I_1}{\text{dim}(\chi_1)}\right)\right].
\end{equation*}
Here, we have used the fact that the average of $\dyad{\psi_0}$ results in the maximally mixed state $I_1/\text{dim}(\chi_1)$. For a system of two transmons, the computational subspace $\chi_1$ is spanned by $\{\ket{00},\ket{01},\ket{10},\ket{11}\}$, and thus $\text{dim}(\chi_1){=}4$. 

Intuitively, the average leakage $L$ quantifies the population fraction initially prepared in the computational subspace, that ultimately ends up outside of  this subspace~\cite{Wood_2018_leakage}. While a prerequisite for achieving a high-fidelity quantum gate is to minimize leakage at gate completion, conventional anti-leakage schemes typically suppress leakage throughout the entire gate duration. This is due to the difficulty in restoring population into the computational subspace at the end, as well as (historically) disproportionately larger decoherence rates for higher levels. Nevertheless, high-fidelity control solutions, with considerable excursion beyond the computational subspace during gate duration, do in fact exist and are achievable with the use of RL optimization, as we will demonstrate later, e.g., see the data reported for RL protocols in Fig.~\ref{fig:shortergatetime} and the corresponding discussion in Sec.~\ref{subsec:shortergate}.

\subsection{Entanglement}\label{subsec:entanglement}
In addition to the fidelity, an important goal of a two-qubit operation is to generate entanglement. Among a number of different options, we select a simple metric called \emph{linear entropy} which quantifies the entanglement of a joint density matrix $\rho$ describing the pure state of both qubit $A$ and qubit $B$ as follows
\begin{equation}
    S_{\text{lin}} = 1 - \Tr_B \left[(\Tr_A\rho)^2\right],
\end{equation}
where $\Tr_A$ ($\Tr_B$) denotes partially tracing out qubit $A$ $(B)$. To calculate the linear entropy of an initial state $\ket{\psi_0}$ after a unitary operation $U$, we simply substitute $\rho=\dyad{\psi}=U\dyad{\psi_0}U^{\dagger}$.  When applied to a multi-level system like transmons, we make sure to normalize the final state after projecting to the qubit subspace. 

To assess the entanglement capacity of a quantum gate, we draw inspiration from the widely adopted \emph{entangling power} of unitary operations~\cite{EntanglingPower_Zanardi_2000}, defined as the average linear entropy produced by a unitary operator when acting on the space of all two-qubit product states. Since the average is taken over two single-qubit Haar measures instead of a single joint two-qubit Haar measure, it can be computed exactly using the set of tensor products of all six Pauli eigenstates $\{\ket{0},\ket{1},\ket{\pm},\ket{\pm i}\}$. Of the resulting 36 two-qubit product states, 16 are maximally entangled, while 20 remain separable for gates in the class of locally equivalent CNOT operations, including the $ZX(\pi/2)$ gate. Within the scope of our work, the linear entropy averaged over these 16 initial states is sufficient to capture the entangling power of a unitary operation. We shall therefore define this quantity as the \emph{average linear entropy} $\bar{S}_{\rm lin}$.

In the context of driving one qubit at a frequency of another in order to implement an entangling gate, it is common to attribute the entanglement generated entirely to the use of such a cross-resonance drive. However, this might not be the case when an on-resonance drive is used simultaneously with the cross-resonance drive. As we shall see, studying the average linear entropy $\bar{S}_{\rm lin}$ of optimized pulses implementing two-qubit gates reveals that in some of the control solutions discovered by RL, the roles of different drives are not as isolated as one might initially believe; e.g., see Fig.~\ref{fig:roles_of_drives}, indicating the existence of an entirely new class of solutions.

\section{State-of-the-art implementations of transmon gates} \label{sec:default_implementation}

\begin{figure}[t!]
    \centering
    \includegraphics[width=0.49\textwidth]{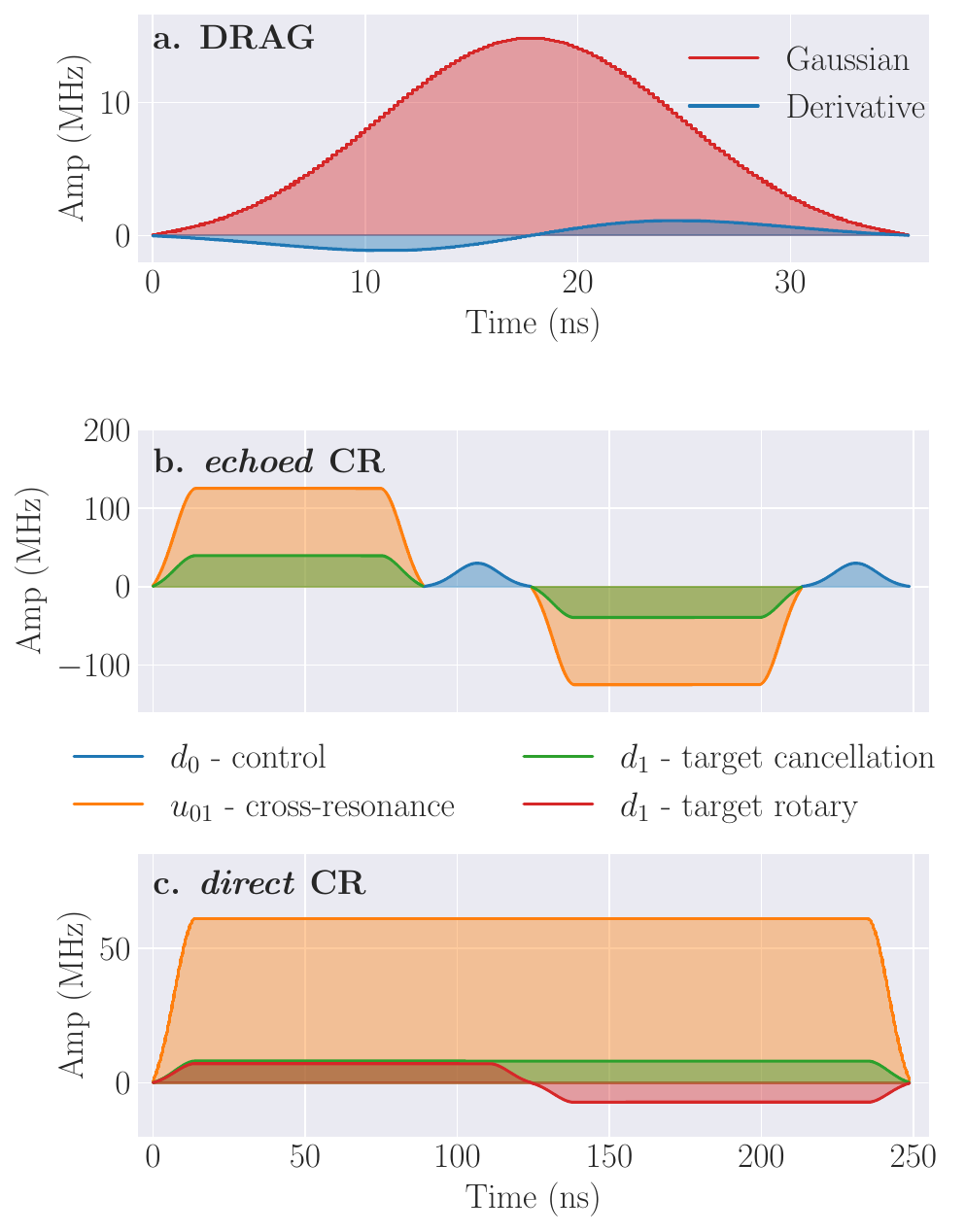}
    \caption{\textbf{Standard error suppression techniques for implementing gates on transmon-qubit devices.} The analytical waveforms are discretized at the inverse sampling rate $dt=2/9$ ns. a) $R_X(\pi/2)$ implemented using DRAG scheme with an in-phase Gaussian pulse $d(t)$ and its out-of-phase derivative (blue). $ZX(\pi/2)$ implemented using b) an \emph{echoed} pulse and c) an echo-free/\emph{direct} pulse, consisting of a main cross-resonance pulse $u_{01}(t)$ (orange) along with on-resonance drives, $d_0(t)$ and $d_1(t)$, on control (blue) and target (green and red) qubits.
    }
    \label{fig:ibm_defaults}
\end{figure}

To provide a meaningful point of comparison for our approach, we review the conventional methods for implementing quantum gates within a superconducting platform and specifically the theoretical foundations underpinning each ansatz. As concrete examples, we showcase the standard error suppression techniques for both the single-qubit $R_X(\pi/2)$ gate and the two-qubit $ZX(\pi/2)$ gate in Fig.~\ref{fig:ibm_defaults}.

The most basic implementation of a single-qubit rotation around the $x$-axis involves driving the target transmon resonantly with a real-valued pulse envelope $d(t)$, according to Eq.~\ref{eq:Hctrl_1}. Common choices for the pulse shape include Gaussian and Gaussian Square waveforms as they offer smooth ramp-up and ramp-down. The standard error suppression approach utilizes an additional out-of-phase component equal to the derivative of the in-phase part, see Fig.~\ref{fig:ibm_defaults}a, which has been shown to significantly reduce gate error including leakage to the second excited level. This is known as the Derivative Removal for Adiabatic Gate (DRAG) scheme \cite{Motzoi_2009, theis2018dragreview}, in which the amplitude of the real Gaussian pulse, the detuning, and the scaling factor of the imaginary derivative component can be optimized. Calibration of these parameters on current superconducting hardware can reliably achieve average gate fidelity of above 99.95\%~\cite{jurcevic2020demonstration}.

For two-qubit entangling gates such as $ZX(\pi/2)$, the standard implementation makes use of a cross-resonance pulse $u_{01}$ along with resonant drives $(d_0,d_1)$ on both the control and target qubit, according to Eq.~\ref{eq:H2total}. These components can be combined in an \emph{echoed} or \emph{direct} fashion~\cite{jurcevic2020demonstration}. As illustrated in Fig.~\ref{fig:ibm_defaults}b, the \textit{echoed} scheme employs an echo pulse sequence where the CR pulse is broken into two halves (yellow envelopes) with the second one inverted $(\Omega\rightarrow-\Omega)$ and positioned between two $\pi$-pulses applied to the control qubit
\begin{equation}
    \text{CR}\left(\Omega\right)\cdot \big[ XI \cdot \text{CR}\left(-\Omega\right) \cdot XI \big].
\end{equation}
The amplitude inversion changes the sign of $\omega_{ZX}$ and $\omega_{IX}$ according to the relations in Eq.~\eqref{eq:interactionrate}, while the addition of two $\pi$-pulses can be understood as a conjugation by $XI$ for every term in the effective CR Hamiltonian, leading to the following contribution from the second half 
\begin{eqnarray}
    \begin{bmatrix}
       ZX\\ IX\\ ZI\\ ZZ
    \end{bmatrix}
    &\xrightarrow{\Omega\rightarrow -\Omega} 
    \begin{bmatrix}
      -ZX\\ -IX\\ ZI\\ ZZ
    \end{bmatrix}
    &\xrightarrow{\text{conj. by }XI} 
    \begin{bmatrix}
      ZX\\ -IX\\ -ZI\\ -ZZ
    \end{bmatrix}.
\end{eqnarray}
When combined with the first half, this should ideally lead to a complete cancellation of unwanted $IX,ZI$, and $ZZ$ terms. Nevertheless, experimental results reveal a significant $IY$ component as well as a small $ZY$ term, which can be attributed to classical crosstalk. This issue can be rectified by applying an on-resonance tone to the target qubit with an identical waveform as the CR pulse, known as active cancellation~\cite{Sheldon_2016}. On the other hand, the \textit{direct} scheme employs an echo-free sequence with the same symmetric active cancellation tone, while introducing an additional asymmetric rotary component. In particular, the symmetric part reduces the effects of $IX$ and $IY$ terms whereas the asymmetric part helps offset $ZZ$ and $ZY$ terms. For both schemes, calibration of the amplitudes and phases of the main CR pulse, in tandem with calibration of the additional tones, achieves between 99\% to 99.7\% average gate fidelity~\cite{Sheldon_2016,jurcevic2020demonstration}.  

As seen in the above examples, the standard pulse designs rely heavily on a theoretical understanding of the platform, i.e., types of interaction induced when certain control drives are activated or when certain error processes are present. On a real device, however, deviation from the theoretical model is unavoidable and closed-loop optimization is required to mitigate the unwanted effects. Additionally, the perturbative approach of deriving the effective interaction rates break down at high control amplitudes, preventing exploration for potential solutions in the strong drive and short time regime. 

While these theoretical ans\"atze offer the advantage of straightforward calibration procedures with a minimal number of parameters, they may also impose significant limitations and/or necessitate longer gate durations to compensate for errors not captured by the relevant theoretical model.  Moreover, should previously unidentified errors come to light, it will be necessary to develop and implement novel error suppression strategies. Established alternative approaches usually involve gradient-based optimization, such as GRAPE~\cite{khaneja2005optimal}, which still requires detailed knowledge about the model and access to the gradient of the loss function. A model-free approach like reinforcement learning is therefore highly desirable since it offers adaptability to system dynamics by learning from ``direct interactions'' which we will define in the next section. Even when equipped with a relatively simple but flexible design space, such as piece-wise constant pulses, RL has the potential to unearth control solutions that are out of reach in conventional methods~\cite{Dalgaard_2020_alphazero}. Furthermore, RL leaves us with a representation of gained knowledge from the control problem, i.e., the agent, that can be reused and analyzed for additional insights (cf.~Sec.~\ref{sec:results}).

\section{Reinforcement learning}\label{sec:rl}

\begin{figure}[t!]
    \centering
    \includegraphics[width=0.3\textwidth]{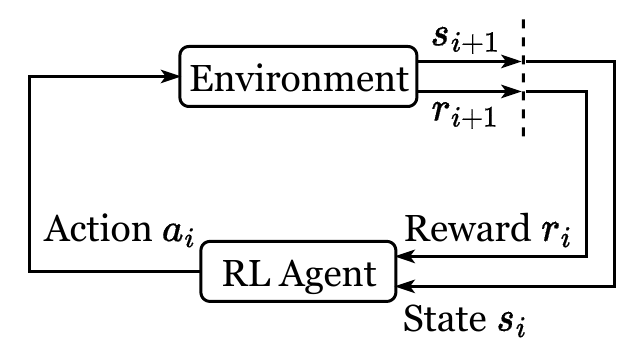}
    \caption{\textbf{Basic reinforcement learning loop.} The agent interacts with its environment (different from the conventional definition of environment in physics) by taking actions and in turn receiving information about the environment's new state. In addition, the agent receives a reward indicating the usefulness of the last action for achieving the given task.}
    \label{fig:simple_rlloop}
\end{figure}

Reinforcement learning operates on a simple principle of trial and error. A generic problem involves an agent learning to make decisions to complete a task by interacting with an environment. Therefore, it is natural to formulate a reinforcement learning problem using a finite Markov Decision Process, in which a decision is made based solely on the current state of the system but not the entire history. We illustrate the basic reinforcement learning loop in Fig.~\ref{fig:simple_rlloop}. In this framework, at every step $i$, the agent selects an action $a_i$ based on a probability distribution or \emph{policy} $\pi(a_i|s_i)$, conditioned on the current state of the environment, $s_i$. After execution, the agent observes a new state $s_{i+1}$ along with a reward $r_{i+1}$ which indicates progress towards completing a particular task. The process terminates once the task is completed or the number of steps reaches a set limit, defining the end of an \emph{episode}. Training the agent then involves running many of these episodes to gather experience, while exploration is encouraged by adding randomness to the action selection procedure, the \emph{trial} process. At the same time, the policy is iteratively adjusted to maximize the expected cumulative reward $\mathbb{E}\left[\sum_i r_{i+1}\right]$ at the end of each episode, guiding the agent away from unproductive actions, the \emph{error} process. Together, trial and error allow the RL agent to explore new actions effectively, and eventually arrive at a highly-rewarded behavior. 

An agent trained exclusively on a single environment typically excels only within that specific context, making it less adaptable when confronted with a new environment. To mitigate this limitation, one effective strategy is to expand the agent's training scope to encompass a variety of environments. Moreover, equipping the agent with some \textit{context} information about its current environment can significantly enhance its learning process and overall decision-making capability. This framework is commonly referred to as reinforcement learning with \textit{context}~\cite{benjamins2022contextualize}, and it has been demonstrated as particularly valuable for tasks that require generalization to a range of environment parameters.

We now adapt the RL framework to designing quantum gates, in which we task an agent to build a piece-wise constant (PWC) pulse to realize a target operation on a transmon environment, as illustrated in Fig.~\ref{fig:rlloop}. In the following, we detail our simulated environment, motivate our choice of states, actions, and rewards, and describe the selected RL algorithm.

\subsection{Environment}

\begin{figure*}[t!]
    \centerline{
        \includegraphics[width=1
\textwidth]{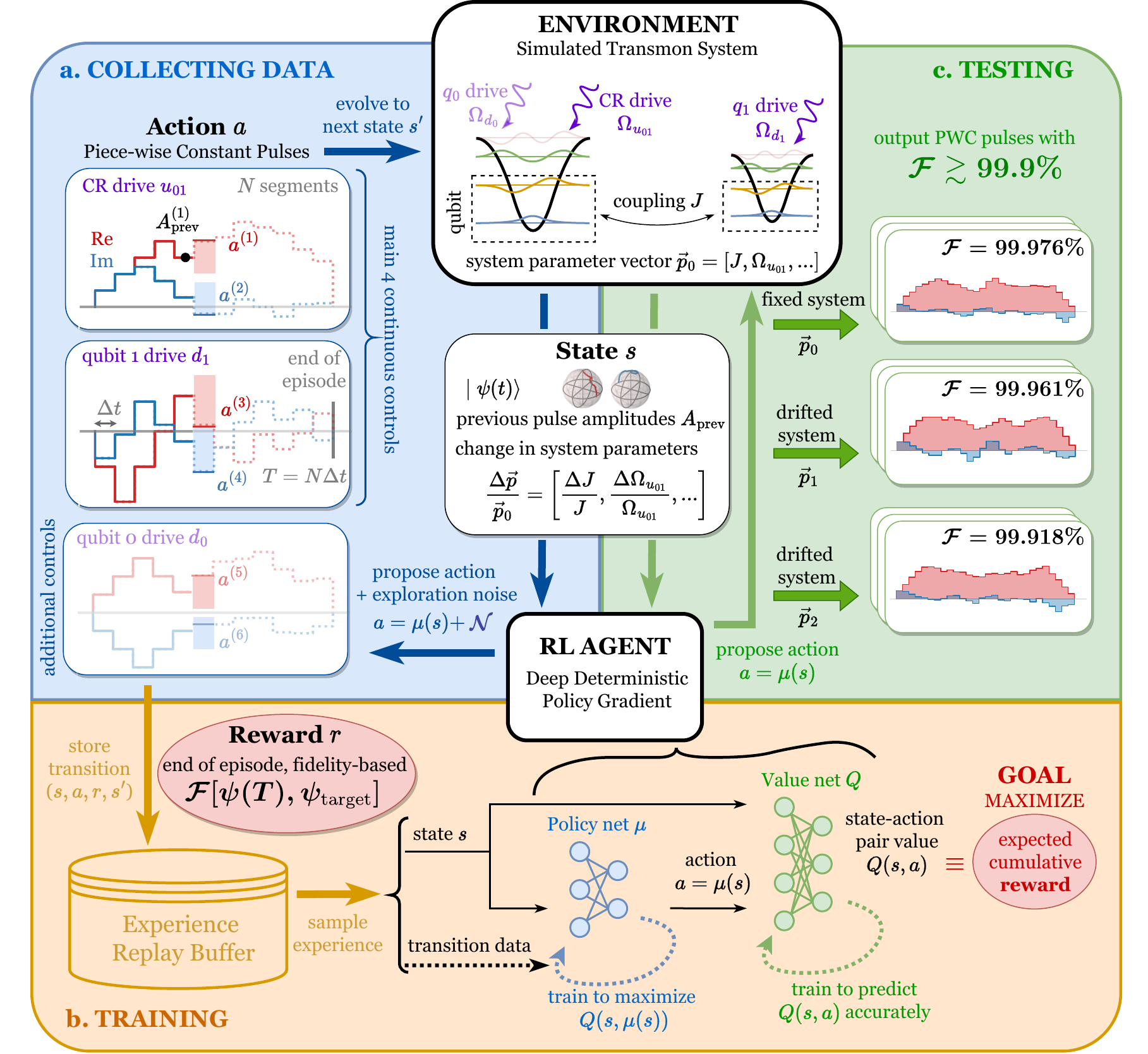}}
          \caption{\textbf{Reinforcement learning for designing high-fidelity quantum gates.} RL framework involves two main entities: the \textit{environment}, which is a system of two coupled transmons simulated as anharmonic oscillators truncated a three energy levels, and the \textit{RL agent}, which uses the DDPG algorithm for learning continuous control drives. We focus on learning 2 control drives (cross-resonance $u_{01}$ and qubit 1 rotation $d_1$) in the main text, and report additional results for including a third control drive (qubit 0 rotation $d_0$) in App.~\ref{app:3drives}.  a) \textit{Step 1: Collecting data.} At every step, the current state $s$ of the environment is characterized by the time-evolved quantum state of the transmons $\{\psi_j(t)\}$, the previous control pulse amplitudes $A_{\rm prev}$, and the relative changes in system parameters $\Delta \vec{p}/\vec{p}_0$. Based on that state $s$, the RL agent proposes an action $a$ to determine control drive amplitudes that evolve the transmon environment forward in time. The environment outputs the next state $s'$ and a fidelity-based reward $r$ (cf.~Eq.~\ref{eq:reward}), and the transition tuple $(s,a,r,s')$ is stored to an Experience Replay Buffer. An episode is complete when the RL agent fully constructs an $N$-segment pulse, and data from many episodes are collected for training. Here we consider a sparse reward scheme, meaning a non-zero reward is given only at the end of each episode. In addition, during data collection, some noise is injected into the RL agent's action to encourage exploration of new control solutions (cf.~Alg.~\ref{alg:ddpg}). \textit{b) Step 2: Training.} Transition data from the Experience Replay Buffer are randomly sampled for batch-training two networks in DDPG algorithm: a \textit{value network} $Q$, which learns to accurately predict the expected cumulative reward $Q(s,a)$ of taking an action $a$ from a state $s$, and a \textit{policy network} $\mu$,  which learns to propose an action $a=\mu(s)$ that maximizes this $Q$-value. Outside of this training process, \textit{RL agent} typically refers to the policy network $\mu$ because it generates all of the agent's actions.  c) \textit{Step 3: Testing.} Once trained, the RL agent can deterministically construct pulses with fidelity $\gtrsim 99.9\%$, not only for a fixed environment, but also for environments whose parameters have drifted. 
         }
    \label{fig:rlloop}
    % \vspace{0.095cm}
\end{figure*}

Our environment simulates the dynamics of two transmons according to the Hamiltonian in Eq.~\ref{eq:H2total}, considering them as directly coupled anharmonic oscillators which can be controlled via external microwave pulses (cf.~``Environment" box in Fig.~\ref{fig:rlloop}). The first two levels of the oscillators act as qubits and the main contribution of leakage out of the qubit subspace is captured via inclusion of the third level. The environment is then completely characterized by a set of system parameters, including detuning $\{\delta_0,\delta_1\}$, anharmonicity $\{\alpha_0,\alpha_1\}$, control drive strength $\{\Omega_{d_0},\Omega_{u_{01}},\Omega_{d_1},\Omega_{u_{10}}\}$, and coupling $J$, which we collect into a single vector $\vec{p}=[J,\Omega_{u_{01}},...]$. We denote the main set of system parameters used in this work as $\vec{p}_0$ whose components are summarized in Table~\ref{tab:deviceparam}. Any drifts in the system characteristics are considered w.r.t to $\vec{p}_0$ via the relative change $\Delta \vec{p}/\vec{p}_0=(\vec{p}-\vec{p_0})/\vec{p}_0$, where we have used an element-wise vector division.

\textit{Action:} The RL agent interacts with the transmon environment by directly modifying the complex-valued control pulses $[u_{01}(t),d_1(t),...]$. Using the PWC ansatz, pulse shaping is equivalent to picking an amplitude $A$ at each discrete time step $\Delta t$ until an $N$-segment pulse is complete, resulting in a gate duration $T=N\Delta t$. To maintain experimental viability and avoid unrealistic oversampling, the time step is chosen such that $1/\Delta t$ be below the sampling rate (and bandwidth) of standard control electronics \cite{Motzoi_2011}, i.e., $\Delta t> dt=2/9$ ns in this work. From the RL point of view, each complete pulse constitutes an episode, after which, the environment is reset so a new pulse can be tried out. Allowing the pulse amplitude $A$ to take any value at every step tends to result in highly volatile pulses, similar to those typically obtained from unconstrained optimal control using methods like GRAPE~\cite{khaneja2005optimal}. Instead, we aim for slowly varying solutions by defining the agent's action to be the relative amplitude change and restrict it to some continuous window $[-w,w]$. By setting $u_{01,i} \equiv u_{01}(i\Delta t)$ and $d_{1,i} \equiv d_{1}(i\Delta t)$, we can write the action at step $i$ as
\begin{eqnarray}
    a_i &=& A_i - A_{i-1} \nonumber \\
    \Rightarrow \begin{bmatrix}
        a_i^{(1)}\\
        a_i^{(2)}\\
    a_i^{(3)}\\
    a_i^{(4)}\\
        \vdots    
    \end{bmatrix}
    &=& \begin{bmatrix}
        \Re(u_{01,i})\\
        \Im(u_{01,i})\\
        \Re(d_{1,i})\\
        \Im(d_{1,i})\\
        \vdots    
    \end{bmatrix}
    - \begin{bmatrix}
        \Re(u_{01,i-1})\\
        \Im(u_{01,i-1})\\
        \Re(d_{1,i-1})\\
        \Im(d_{1,i-1})\\
        \vdots    
    \end{bmatrix}, \label{eq:action}
\end{eqnarray}
where each component is restricted to a drive-dependent window
\begin{equation}
    \abs{a_i^{(1)}}, \abs{a_i^{(2)}} \leq w_u, \qquad
    \abs{a_i^{(3)}}, \abs{a_i^{(4)}} \leq w_d. 
\end{equation}
Hence, the action space dimension corresponds to twice the number of available control fields, as shown in Fig.~\ref{fig:rlloop}a. By choosing the windows $w_u$ and $w_d$ to be small, e.g., less than 10\% of the maximum allowed range, we systematically restrict the action space which additionally improves learning. Finally, we clip the resultant amplitudes to [-1,1] to ensure that the control fields do not exceed the maximum allowed drive strengths.

\textit{State:} The evolution of the transmon system due to external control fields and internal dynamics is characterized by a unitary map $U(t,0)$ computed from Hamiltonian in Eq.~\ref{eq:H2total}. Given a set of basis states $\{\psi_j\}$, the evolution of an arbitrary pure initial state reads
\begin{equation}
    U(t,0)\ket{\psi} = \sum_j c_j U(t,0)\ket{\psi_j} = \sum_j c_j \ket{\psi_j(t)}.
\end{equation}
Thus, tracking the time-evolved unitary map is equivalent to tracking the time-evolved basis states $\{\psi_j(t)\}$. As we aim to design target operations between two-level systems, we assume no occupation beyond the qubit subspace initially. That means, for example, in the single-qubit gate case, it is sufficient to track only the following basis states
\begin{eqnarray}
    \ket 0 = (1,0,0)^T,\quad \ket 1 = (0,1,0)^T,
\end{eqnarray}
\clearpage

\noindent
where we have truncated our simulation at three levels. Complete knowledge of the evolved basis states $\{\psi_j(t)\}$ at every step allows the agent to discern the effect of its actions on the environment. Due to our restriction of the action space to contain only relative changes in the control fields as in Eq.~\ref{eq:action}, we also include the pulse amplitudes in the previous time step to the state provided to the agent:
\begin{equation}
    s_i = [\{\psi_j(t=i\Delta t)\},A_{i-1}]. \label{eq:state1}
\end{equation}
In addition to designing gates for a fixed environment, we also wish to generalize the agent's designing capability to environments where the system parameters have drifted from their original values. While the agent can indirectly discern this change through the evolution of the quantum state, we have observed that furnishing the agent with explicit information about the current system characteristics can enhance its learning process. Instead of directly feeding the agent the system parameter vector $\vec{p}$ whose entries can take a wide range of values, we can provide the same \emph{context} information via the relative change in system parameters $\Delta \vec{p}/\vec{p}_0$, transforming the RL input state into
\begin{eqnarray}
    &s_i \leftarrow \left[s_i,\frac{\Delta \vec{p}}{\vec{p}_0}\right],\label{eq:state2}
\end{eqnarray}
where $\vec{p}_0$ denotes the original values in Table~\ref{tab:deviceparam}.

\textit{Reward:} RL approaches typically utilize a reward that is provided at every step to incentivize the agent to learn the correct actions. Alternatively, the agent can also learn from a single reward granted at the end of each episode. In the case of a fidelity-based objective and when considering a closed-loop implementation using an actual device, this sparse reward scheme demands fewer measurements during intermediate steps, making it more experimentally favorable. With this in mind, we have opted for the sparse reward scheme and have determined that it is adequate for the agent's learning process.  As the fidelity approaches unity, improvements tend to slow down, yielding increasingly marginal gains. To enhance the discernibility of positive signals, we define the reward function to be the negative log infidelity at the final time step
\begin{eqnarray}
r_N = -\log_{10}(1-\mathcal{F}). \label{eq:reward}
\end{eqnarray}
Here, it is important to note that an improvement of one unit in the reward corresponds to a one-order-of-magnitude enhancement in fidelity, e.g., $r:2\rightarrow3$ corresponds to $\mathcal{F}:0.99\rightarrow0.999$.

\subsection{Algorithm}

In our pursuit of designing quantum gates via pulse shaping, we have established a large design space of PWC functions for the RL agent to explore. We require an algorithm capable of handling continuous-valued actions to fully harness the flexibility of this design space for achieving high-fidelity solutions, and also to effectively utilize continuous control resources in realistic hardware. Furthermore, given the limited access to near-term quantum devices, an algorithm with efficient training data usage is highly desirable. Thus, we select the\ Deep Deterministic Policy Gradient algorithm (DDPG), which satisfies all of these criteria~\cite{lillicrap2019continuous}.

We begin by laying the groundwork for DDPG, which is rooted in the concept of Q-learning. A Q-value quantifies the expected cumulative reward associated with taking an action from a specific state and subsequently following a particular policy $\pi$ thereafter. In reinforcement learning, the expected cumulative reward is commonly subject to discounting over future time steps in order to incentivize the agent to complete its objective faster to receive a higher reward. Formally, the Q-value for a state-action pair $(s_i, a_i)$ at time step $i$ under a policy $\pi(a|s)$ is defined as follows:
\begin{eqnarray}
    Q^{\pi}(s_i,a_i) &=& \mathbb{E}_{\pi}\left[r_{i+1} +  \gamma r_{i+2} + \dots|s_i,a_i\right]  \nonumber \\
    &=& r_{i+1} +  \gamma\mathbb{E}_{\pi}\left[r_{i+2} + \dots\right] \nonumber \\
    &=& r_{i+1} +  \gamma\sum_{a_{i+1}} \pi(a_{i+1}|s_{i+1})Q^\pi(s_{i+1},a_{i+1}). \nonumber \\
\end{eqnarray}
Here,  $\gamma\in[0,1]$ is the discount factor, and the expectation value $\mathbb{E}$ is taken over actions selected using the policy $\pi$. With the current action $a_i$ already selected, there is only one possible value for the immediate reward $r_{i+1} = r_{i+1}(s_i,a_i)$, allowing us to take it out of the expectation and substitute it in the Q-value definition for the next state-action pair. The optimal strategy is then to pick the highest-valued action at every step, which leads to the recursion relation for the optimal Q-value $Q^*$:
\begin{eqnarray} \label{eq:Bellman}
    Q^*(s_i,a_i) &=& \max_{\pi} Q^{\pi}(s_i,a_i)  \nonumber\\  
    &=& r_{i+1} + \gamma\max_a Q^*(s_{i+1},a),
\end{eqnarray}
also known as the Bellman optimality equation \cite{suttonbarto}. As the dependence on the policy $\pi$ is removed in the above, the optimal Q-value can be iteratively updated using any transition data tuple $(s_i,a_i,s_{i+1},r_{i+1})$ regardless of the collecting policy, a process commonly known as off-policy training. In practice, observed transitions are stored in a replay buffer from which a mixture of new and old transitions are sampled to train the agent. A typical replay buffer stores about half a million transitions, allowing much more efficient re-use of old data as compared to other RL methods.

When the number of discrete states and discrete actions are not too large, the corresponding Q-values are stored in a finite-size table that can be iteratively updated. As the state space becomes continuous, one instead approximates the optimal Q-value by a deep neural network with parameters $\phi$ as $Q_\phi \approx Q^*$. To adapt this deep Q-learning method to continuous actions, DDPG additionally utilizes a deterministic policy network with parameters $\theta$ for action generation: $a_i=\mu_{\theta}(s_i)$. During training, a noise process is injected into the agent's action to encourage exploration as seen in Fig.~\ref{fig:rlloop}a. With the main goal of maximizing the expected cumulative reward, we want not only a policy network that can generate actions with high Q-values, but also a value network that can approximate the Q-values well according to Eq.~\ref{eq:Bellman}, which leads to the following update rules:
\begin{eqnarray}
    Q_\phi(s_i,a_i) &\leftarrow& r_{i+1} + \gamma Q_{\phi}(s_{i+1},\mu_\theta(s_i)), \nonumber \\
    \mu_\theta (s_i) &\leftarrow& \arg\max_{a_i} Q_\phi(s_i,a_i), \label{eq:updaterules}
\end{eqnarray}
for each transition data tuple $(s_i,a_i,s_{i+1},r_{i+1})$. From these update rules, it is clear that updating one network changes the loss function of the other, creating a moving target problem. Therefore, when computing the targets on the right hand sides of Eqs.~\ref{eq:updaterules}, we employ target networks $(\phi',\theta')$ that slowly track the learned networks $(\phi,\theta)$, which minimizes the effect of fast-moving targets. The complete algorithm is detailed in App.~\ref{app:ddpg}.

DDPG is known to struggle when the action space gets too large, which leads to exploding Q-values during training. One solution is to train two Q-networks and use the smaller value for computing the targets in Eq.~\ref{eq:updaterules} to mitigate Q-value overestimation (a.k.a., twin network trick). Another solution is to delay the policy network update for better Q-network convergence in between (a.k.a. delayed policy trick). These techniques, when combined with the standard DDPG, result in an augmented algorithm commonly known as Twin Delayed DDPG (TD3) \cite{TD3_fujimoto2018addressing}. Unfortunately, we could not obtain conclusive evidence as to whether TD3 outperforms DDPG in all situations for our problem. Therefore, in the main text we focus on DDPG and delegate discussion of a case in which TD3 provides an advantage to App.~\ref{app:3drives}.

\section{Results} \label{sec:results}

\begin{figure}[t!]
    \centering
    \includegraphics[width=0.48\textwidth]{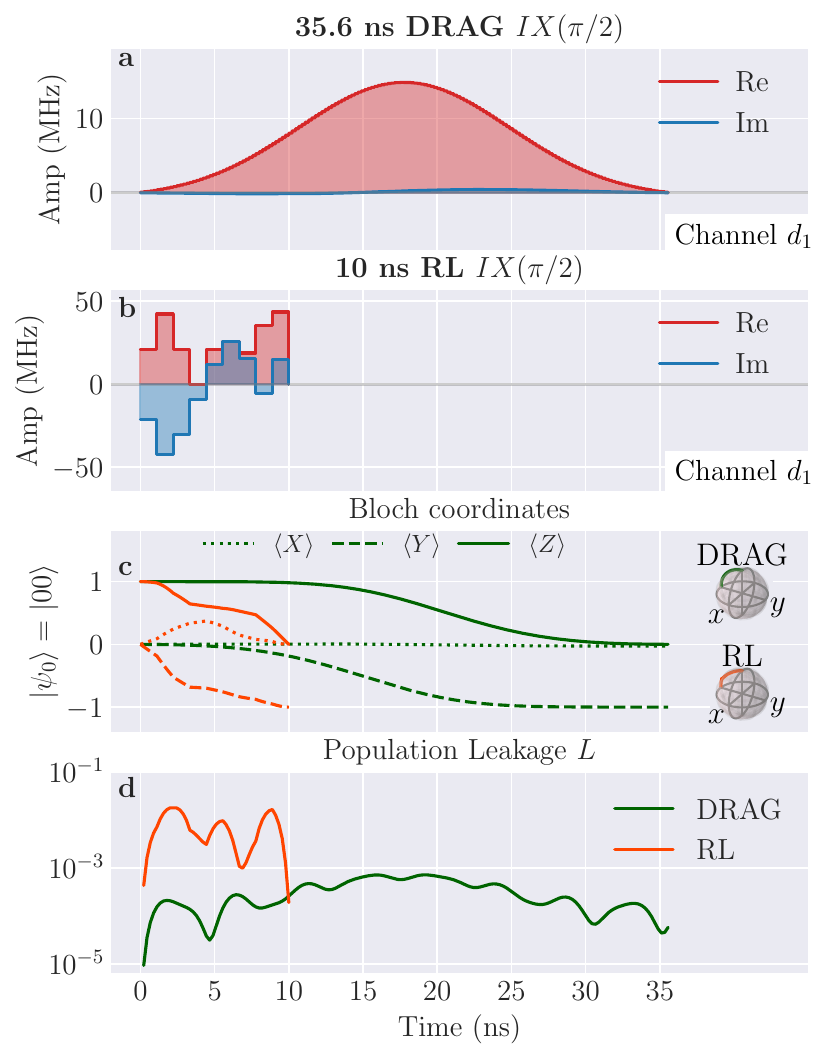}
    \caption{\textbf{Optimization for the single-qubit gate $X(\pi/2)$ in two-transmon setting.}
    (a) IBM 35.6 ns DRAG pulse. 
    (b) 10 ns RL-optimized pulse with training hyperparameters given in the ``Fixed Environment" section of Table~\ref{tab:hyperparameters}.
    (c) Corresponding evolution of Bloch coordinates for the controlled qubit. 
    (d) Population leakage to $\ket{2}$ for RL pulse is up to a few orders of magnitude higher compared to DRAG during the evolution. 
    RL-optimized pulse is $3\times$ faster with similar average gate fidelity above 99.9\%, and makes use of the presence of level $\ket{2}$, at the expense of accessing three times larger amplitudes.
    }
    \label{fig:learnfromscratch_1q}
\end{figure}

\begin{figure*}[t!]
    \centerline{
        \includegraphics[width=1\textwidth]{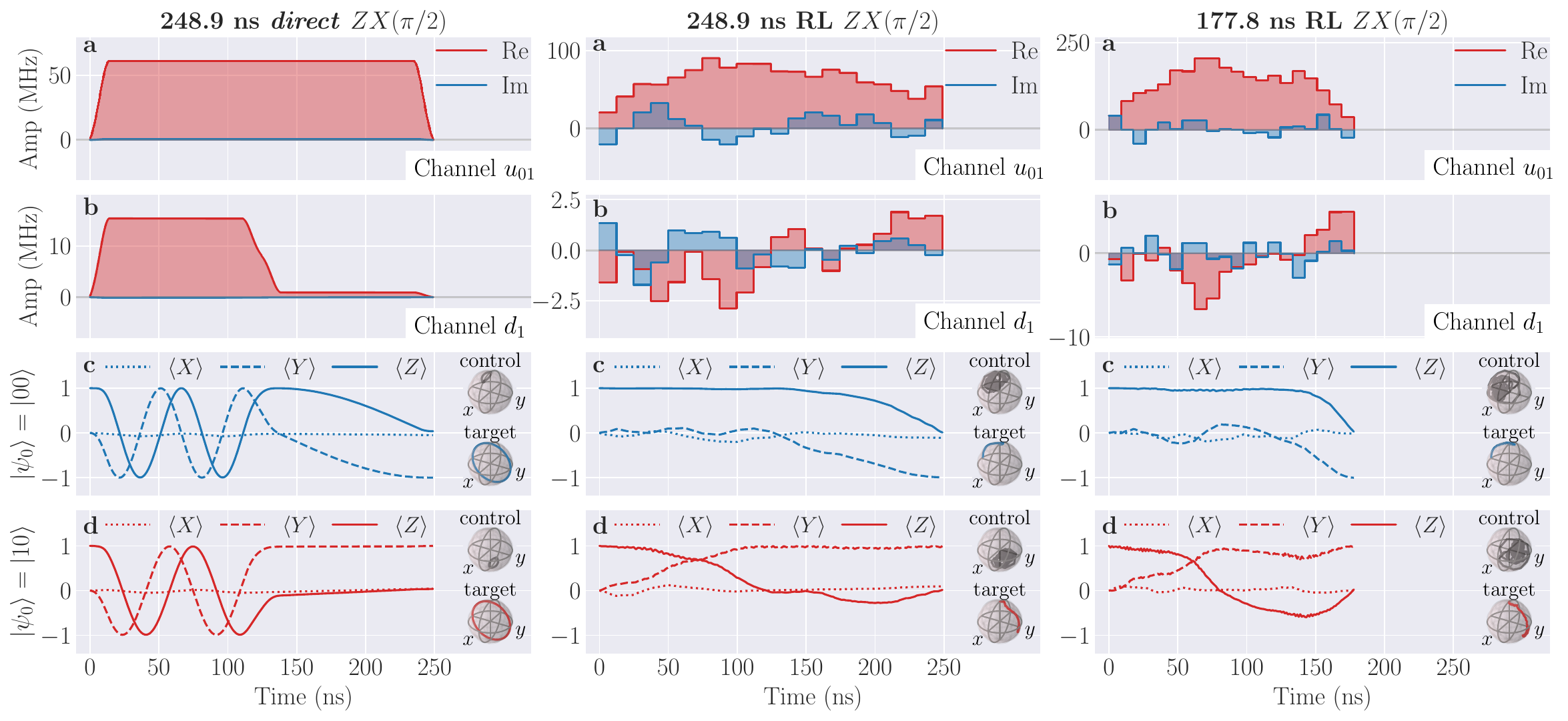}}
        \caption{\textbf{Optimization for the cross-resonance gate $ZX(\pi/2)$.} We display results with fidelity over 99.9\% for the \emph{direct} and RL approaches at gate durations 248.9 ns and 177.8 ns. 
        (a-b) Optimized pulse envelopes for the cross-resonance drive $u_{01}$ and target qubit drive $d_1$. (c-d) Corresponding evolution of Bloch coordinates for target qubit when the control state is $\ket 0$ or $\ket 1$.  Pulses designed by our RL agent appear considerably different from the \emph{direct} scheme, in both pulse shape and quantum state dynamics. Furthermore, our RL agent manages to shorten the gate duration to 177.8 ns without compromising $99.9\%$ fidelity.
        RL training hyperparameters are given in the ``Fixed Environment" section of Table~\ref{tab:hyperparameters}.
        }
        \label{fig:learnfromscratch_2q}
\end{figure*}

Here we report our main results of utilizing the DDPG algorithm for continuous control to solve the two-qubit gate design problem for superconducting transmon qubits. To ensure the consistency and reproducibility of the RL approach, we repeat each case multiple times with different seeds. Each seed leads to a different set of random generators used for initializing and optimizing of neural network parameters, sampling experience from the replay buffer, and injecting exploration noise to agent action during training. Unless stated otherwise, the reported results appear typical within a handful of realizations. The best cases are discussed here in the main text while training data over multiple seeds are included in App.~\ref{app:trainingprocedure}. We summarize relevant training hyperparameters in Table~\ref{tab:hyperparameters} as well as the main set of system parameters in our quantum simulator in Table~\ref{tab:deviceparam}.

We juxtapose the RL-designed strategies with the conventional error suppression schemes in Sec.~\ref{sec:default_implementation} where we employ the Nelder-Mead method to optimize the relevant parameters in each ansatz to maximize the average gate fidelity. For the single-qubit DRAG scheme, we simultaneously optimize 2 parameters for $d(t)$: the amplitude of the real Gaussian pulse and the scaling factor of the imaginary derivative pulse. For two-qubit entangling gates, we find that first optimizing the amplitudes of the Gaussian Square pulses for $u_{01}(t)$ and $d_1(t)$, and then optimizing for their phases, yields the best result. In particular, the \emph{echoed} scheme consists of two tunable Gaussian Square pulses, cross-resonance $(u_{01})$ and target cancellation tone (symmetric part of $d_1$), constituting a 4-parameter optimization problem. Meanwhile, the \emph{direct} scheme contains an additional target rotary tone (asymmetric part of $d_1$), increasing the number of optimizable parameters to 6. These are in contrast with the 18-dimensional (single-qubit gate) and 80-dimensional (two-qubit gate) control problems addressed by our RL approach that we will see shortly.

Our main results can be summarized as follows. First, we demonstrate that an RL agent can be trained via direct interaction with a simulated environment to successfully explore the vast design space of PWC functions (Sec.~\ref{subsec:learnfromscratch}). Although the environment is treated as a black box, the PWC time step is chosen such that $1/\Delta t$ is below the sampling rate (and bandwidth) of standard control electronics \cite{Motzoi_2011}, to avoid unrealistic oversampling.  The discovered strategies are unbiased by prior theoretical knowledge and distinct from the established analytical solutions in Sec.~\ref{sec:default_implementation}. Second, we illustrate the benefit of RL optimization with the flexible PWC ansatz in finding high-fidelity control solutions at shorter gate duration  (Sec.~\ref{subsec:shortergate}). We then assess the novelty in the roles played by each drive (Sec.~\ref{subsec:noveltyinroles}), followed by the robustness of optimized pulses to short-timescale stochastic noise  (Sec.~\ref{subsec:robustness}). Finally, we augment our agent to generalize and adapt to drifts in system characteristics  (Sec.~\ref{subsec:adaptetosystemdrift}), making use of the left-over representation of gained knowledge which is an advantage of RL over conventional control algorithms.

\subsection{Learning without prior knowledge}\label{subsec:learnfromscratch}
\subsubsection{Single-qubit gate in two-transmon setting}

As a benchmark for our approach, we start by tasking the RL agent to learn the single-qubit $\pi/2$ rotation around the $x$-axis in a two-transmon setting, given by
\begin{equation}
    IX(\pi/2) \equiv I\otimes R_X(\pi/2)= I\otimes\left(\frac{I-iX}{\sqrt{2}}\right).
\end{equation}
In Fig.~\ref{fig:learnfromscratch_1q}, we report an RL-designed 9-segment pulse that implements the $IX(\pi/2)$ gate using a single control drive $d_1$. The agent learns to construct both real and imaginary parts of the pulse, thus tackling an 18-dimensional optimization problem. The RL-designed pulse achieves a 10 ns gate duration which is over $3\times$ faster than the 35.6 ns DRAG pulse. With triple the maximum pulse amplitude, the 10 ns pulse maintains a comparable fidelity at 99.9\%, despite having leakage larger by a few orders of magnitude during intermediate steps, as seen in Fig.~\ref{fig:learnfromscratch_1q}d. The data, therefore, suggest that the RL agent learns to exploit the presence of the second level to the advantage of reducing the gate duration. Such speed-up already offers a significant reduction in operating time as general quantum circuits consist mostly of single-qubit gates.

\subsubsection{Two-qubit entangling gates}

Applying the same algorithm for 20-segment waveforms, our RL agent successfully learns 248.9 ns pulses that implement $ZX(\pi/2)$ and CNOT gates completely from the ground up, achieving fidelity $\mathcal{F}>99.9\%$. The agent constructs complex-valued pulses for a cross-resonance drive $u_{01}$ and an on-resonance drive on the target transmon $d_1$, constituting an optimization problem of dimension $80$ ($20\text{ segments}\times 2\text{ drives} \times 2 \text{ real numbers}$). In fact, our agent also finds equally high-fidelity solutions to an even higher-dimensional optimization problem when given access to three drives $(d_0,u_{01},d_1)$. However, these 3-drive control solutions require the ability to send two pulses at different frequencies simultaneously to the same transmon, which, to the best of our knowledge, is not a commonly used technique. Therefore, we defer the discussion of 3-drive results to App.~\ref{app:3drives} and focus on constructing pulses using the only two drives $(u_{01},d_1)$ for our main results section. 

In Fig.~\ref{fig:learnfromscratch_2q}, we present a clear-cut contrast between state-of-the-art \textit{direct} scheme and RL approach for the $ZX(\pi/2)$ gate. First, the RL pulse envelope goes beyond the square Gaussian structure in the \textit{direct} scheme while having a higher maximum amplitude and a more prominent imaginary part, as seen in Fig.~\ref{fig:learnfromscratch_2q}a-b. In Fig.~\ref{fig:learnfromscratch_2q}c-d, we illustrate the evolution of Bloch coordinates of the target qubit initialized at $\ket 0$, for two cases: when the initial control qubit state is either $\ket 0$ (blue) or $\ket 1$ (red). Maximal entanglement is achieved when these two time-evolved target qubit states are exactly opposite on the Bloch sphere. In the \textit{direct} scheme, the conditioned target qubit states start out by rotating together around the $x$-axis at slightly different speeds. They increase their distance after a couple of revolutions and stop at their final destinations on each end of the $y$-axis. By contrast, the RL scheme appears to bring the states directly to their respective destinations with minor course corrections in between; this has to do with our choice of allowed action windows $w_u$ and $w_d$ between the $u_{01}$ and $d_1$ drives. We find that setting $w_u=10w_d$, in this case, yields the best training performance, which inadvertently restricts our agent to solutions with a considerably weaker drive on the target qubit.  More interestingly, the evolution roughly splits into two parts (cf. middle and right columns of Fig.~\ref{fig:learnfromscratch_2q}c-d): the state conditioned on 1 (red) moves while the state conditioned on 0 (blue) remains approximately stationary in the first half of the pulse sequence, and vice versa in the second. The above observations suggest that the strategies learned by our RL agent are fundamentally different from the standard analytical protocols.

\subsection{Achieving shorter gate duration} \label{subsec:shortergate}

\begin{figure}[t!]
    \centering
    \includegraphics[width=0.483\textwidth]{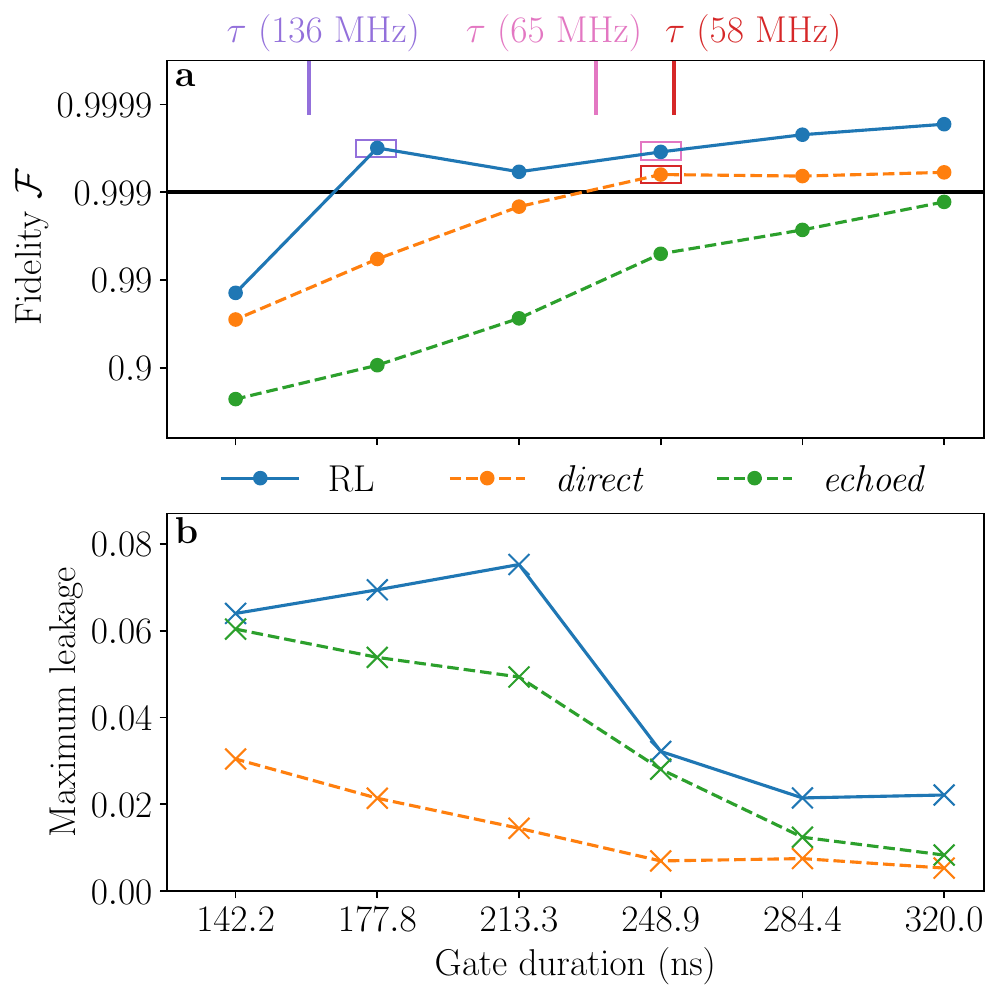}
    \caption{
    \textbf{Fidelity and leakage of optimized $ZX(\pi/2)$ pulses at different gate durations.} We compare results for the \emph{direct}, \emph{echoed}, and RL schemes. 
    (a) Best fidelity achieved over a dozen runs as function of gate duration. Short vertical lines mark the approximate entangling times obtained via numerical block-diagonalization for constant pulses of average amplitudes of the three data points marked by the color boxes. 
    (b) Maximum population leakage throughout gate duration for the same set of pulses. For full evolution of population leakage throughout gate duration, see Fig.~\ref{fig:leakage_gatetimes}. The increase in maximum population leakage coincides with the decrease in fidelity for the \textit{direct} and \textit{echoed} schemes at shorter gate durations. 
    RL-designed pulses, however, maintain 99.9\% fidelity down to 177.8 ns gate duration, potentially making use of large population leakage.
    RL training hyperparameters are given in the ``Fixed Environment" section of Table~\ref{tab:hyperparameters}.
    }
    \label{fig:shortergatetime}
\end{figure}

\begin{figure*}[t!]
    \centerline{
        \includegraphics[width=1\textwidth]{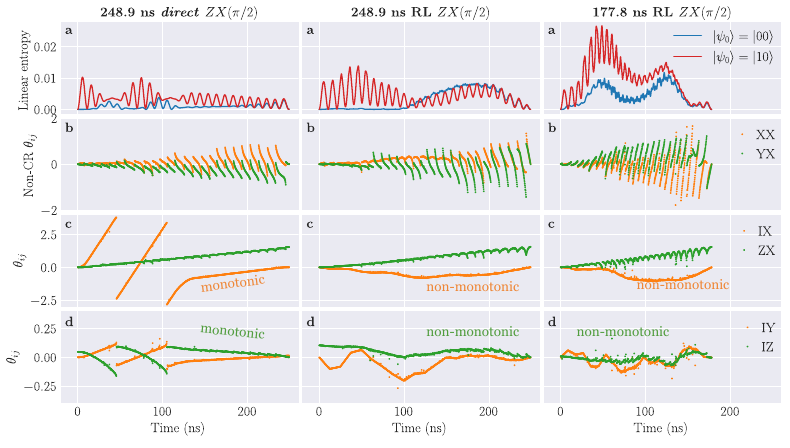}}
        \caption{\textbf{Entanglement and rotation angles of optimized $ZX(\pi/2)$ pulses.} We display results with fidelity over 99.9\% for the \emph{direct} and RL approaches at gate durations 248.9 ns and 177.8 ns.
        (a) Linear entropy $S_{\text{lin}}$ evolution for the initial states $\ket{00}$ and $\ket{10}$, which are expected to remain separable.  
        (b) Rotation angles evolution, cf.~Eq.~\ref{eq:rotangle}, for two entangling interactions \emph{not} present in the effective CR model given in Eq.~\ref{eq:CRHamiltonian}. 
        (c-d) Rotation angles evolution for several interaction types predicted by the effective CR model, including the main entangling process $ZX$ and target qubit rotation angles $IX$, $IY$, and $IZ$. Deviation from zero entanglement for initial states $\ket{00}$ and $\ket{10}$ indicates the presence of entangling interactions beyond this model, such as the shown evolution of $XX$ and $YX$ interactions. While the $ZX$ angle is accumulated in a similar manner, target qubit rotations in RL-designed solutions behave fundamentally differently w.r.t the monotonicity of those rotation angles in the \textit{direct} scheme.}
        \label{fig:entanglement_rotationangles}
\end{figure*}

By extending our gate design study to different gate durations, we find that the RL approach coupled with the flexibility of the PWC waveform consistently outperforms optimized \textit{direct} and \textit{echoed} schemes. In Fig.~\ref{fig:shortergatetime}a, we observe that fidelities obtainable using these standard approaches drop below 99.9\% when their gate durations approach 213 ns and 320 ns, respectively. Meanwhile, RL pulse duration can be shortened significantly, down to 177.8 ns while maintaining the same performance. 

We first compare the gate durations of optimized control solutions with the \emph{approximate entangling time} $\tau$ defined for a constant pulse of amplitude $\Omega$. For the $ZX(\pi/2)$ gate, we have $\tau(\Omega) = (\pi/2)/\omega_{ZX}(\Omega)$, where $\omega_{ZX}$ is the effective $ZX$ interaction rate obtained via numerical block-diagonalization of the two-transmon Hamiltonian in Eq.~\ref{eq:H2CR} (cf.~Ref~\cite{Magesan_2020} for more details). We compute $\tau(\Omega)$ for the three cases considered in Fig.~\ref{fig:learnfromscratch_2q} using their average amplitudes, which are 58 MHz, 65 MHz, and 136 MHz, respectively. These approximate entangling times are displayed as short vertical lines in Fig.~\ref{fig:shortergatetime}a and are color-coded to the corresponding points on the graph. On one hand, we observe that $\tau$ practically equals the gate duration for the 248.9 ns \emph{direct} pulse, which is not surprising as its Gaussian Square waveform can be well-approximated with a constant pulse. On the other hand, for the RL pulses, the gate durations and approximate entangling times no longer agree, which can be attributed to their considerably more complicated PWC waveform. This observation suggests non-trivial dynamics in the control solutions discovered by our RL agent, which we further investigate in the following by examining the amount of population leakage as well as the evolution of entanglement generated in several initial states and rotation angles.

While the target operation involves only the first two levels, they are not isolated from other excited states and input control fields inevitably drive some population out of the computational subspace. In Fig.~\ref{fig:shortergatetime}b, we show the maximum leakage at different gate durations; and observe that an increase in leakage for \textit{direct} and \textit{echoed} pulses coincides with a decrease in fidelity. By contrast, RL-designed pulses manage to preserve their performance despite experiencing large state leakage which inevitably arises as the agent explores high-amplitude solutions in order to shorten the necessary entangling time. Our results suggest a prominent presence of leakage processes beyond the effective model, and that our RL agent has found a way to make good use of them. Indeed, such behavior is supported by prior research which indicates that the improved gate speed can be attributed to the more strongly coupled higher energy levels.~\cite{speedlimit_ashhab_2022}.

Within the computational subspace, we first note the deviation from the effective model by looking at the entanglement generated in two initial states, $\ket{00}$ and $\ket{10}$; they are expected to remain separable throughout the evolution under the effective CR Hamiltonian given in Eq.~\ref{eq:CRHamiltonian}. We show the evolution of the linear entropy $S_{\rm lin}$ of these states in Fig.~\ref{fig:entanglement_rotationangles}a for both \emph{direct} and RL schemes, and observe small but non-zero amounts of entanglement. For the former, we observe more entanglement generated in the $\ket{10}$ state which can be attributed to pulse ramp up and ramp down not taken in account in the effective Hamiltonian analysis~\cite{Magesan_2020}.  For RL-designed pulses, on the other hand, especially one with shorter gate duration, both states become considerably more entangled at intermediate time steps,  indicating the presence of entangling processes beyond those expected from the effective CR model. This suggests that our RL agent has managed to remove these unwanted entangling processes at gate completion in order to achieve a high final fidelity. 

For a more detailed picture of both single-qubit rotation and entangling processes, we take a closer look at the strengths of different interactions in the Pauli basis as a function of time. To do so, we first compute the averaged Hamiltonian by taking the logarithm of the unitary $U(t,0)$ at time $t$ and project it onto the qubit-subspace
\begin{eqnarray}
    tH_{\rm avg}^{\rm qubit} = \Pi^{\rm qubit} [i\ln U(t,0)]\Pi^{\rm qubit}.
\end{eqnarray}
We expand $tH_{\rm avg}^{\rm qubit}$ in the Pauli basis $P_i\in\{I,X,Y,Z\}$
\begin{eqnarray}
    tH_{\rm avg}^{\rm qubit} = \sum_{ij}\theta_{ij}\frac{P_i\otimes P_j}{2},
\end{eqnarray}
where $\theta_{ij}$ defines the rotation angle that depends on the $P_i\otimes P_j$ interaction strength and duration $t$. We can then invert the relation and compute the rotation angle as
\begin{equation}
    \theta_{ij}(t) = \Tr\left[\Pi^{\rm qubit} [i\ln U(t,0)]\Pi^{\rm qubit}\frac{P_i\otimes P_j}{2}\right]. 
    \label{eq:rotangle}
\end{equation}
Computing $\ln U(t,0)$ amounts to choosing an appropriate branch cut to obtain sensible results for the time-dependent rotation angles, a procedure we discuss in App.~\ref{app:rotation_angles}. It is important to note that these branch cuts are chosen to reveal a semi-stable increase in $\theta_{ZX}$, which in turn, results in clear discontinuities in other angles. We can now analyze these rotation angles $\theta_{ij}$ to reveal the arisen interactions in greater depth.

In Fig.~\ref{fig:entanglement_rotationangles}b, we show the time evolution of the rotation angles for the $XX$ and $YX$ interactions. Their non-negligible presence, even in the \emph{direct} scheme, is not expected from the effective CR model given in Eq.~\ref{eq:CRHamiltonian}, which further supports our previous observation of the linear entropy in Fig.~\ref{fig:entanglement_rotationangles}a. The saw-tooth time-evolution of these interactions corresponds to the off-resonance precession of the control qubit, and this pattern differs for all three cases presented. It is worth noting that the short 177.8 ns RL pulse results in the largest precession rate as well as the most prominent $XX$ and $YX$ interactions, which we attribute to its significantly higher drive amplitude as seen in Fig.~\ref{fig:learnfromscratch_2q}. 
We thus confirm that the more complex waveforms in RL pulses feature an increased presence of entangling processes beyond the $ZX$ term.

In Fig.~\ref{fig:entanglement_rotationangles}c-d, we also display several other interactions that are expected from the effective CR model, including $ZX$, $IX$, $IY$, and $IZ$. First, we observe a similar gradual accumulation of $\theta_{ZX}$ across the board. The target qubit rotations, however, look completely different. In the \emph{direct} scheme, an active cancellation tone was introduced to mainly suppress a large unwanted $IX$ term generated by the bare CR pulse. The inclusion of the asymmetric rotary part provides the additional degrees of freedom to suppress more unwanted terms. While having some success, these techniques are limited by the rigidity of the Gaussian Square waveform. This can be seen from \emph{monotonic} evolution of the $IX$, $IY$, and $IZ$ rotation angles in the \emph{direct} pulse as illustrated in the left column of Fig.~\ref{fig:entanglement_rotationangles}c-d. 
By contrast, the evolution of these rotation angles in RL pulses, shown in the middle and right columns of Fig.~\ref{fig:entanglement_rotationangles}c-d, appears to be non-monotonic and significantly more flexible throughout the gate duration, suggesting a much more powerful error suppression potential resulting from the PWC waveform. Indeed, we observe that our RL agent successfully brings all unwanted interactions [cf.~Fig.~\ref{fig:remaining_rotationangles}] for the remaining terms close to zero at gate completion, even in the case of a significantly higher-amplitude 177.8 ns pulse where the effective CR model completely breaks down. These findings reveal a considerable deviation in environment characteristics beyond the perturbative effective dynamical model, and yet, our PWC-equipped model-free RL agent remains unbiased and adjusts accordingly to obtain high-fidelity control solutions.

With the training setting used above, i.e., learning pulses for only 2 drives $(u_{01},d_1)$ via a standard DDPG algorithm, our RL agent only manages to find one 99.9\% fidelity 177.8 ns solution out of a dozen runs. This is because we need to increase the allowed relative change in pulse amplitudes at each step, effectively broadening the action space, in order to compensate for such a short gate duration. Larger action space tends to result in Q-value overestimation, which ultimately destabilizes the training process. By employing a few additional modifications to the training setting such as expanding the agent's access to 3 drives $(d_0, u_{0}, d_1)$, implementing the TD3 tricks, and training the agent for longer, we notice more stable training and an improved probability of success on some occasions but not universally. Therefore, we postpone the discussion of these additional results to App.~\ref{app:3drives}.

\subsection{Novelty in the roles of drives }\label{subsec:noveltyinroles}

\begin{figure}[t!]
    \centering
    \includegraphics[width=0.48\textwidth]{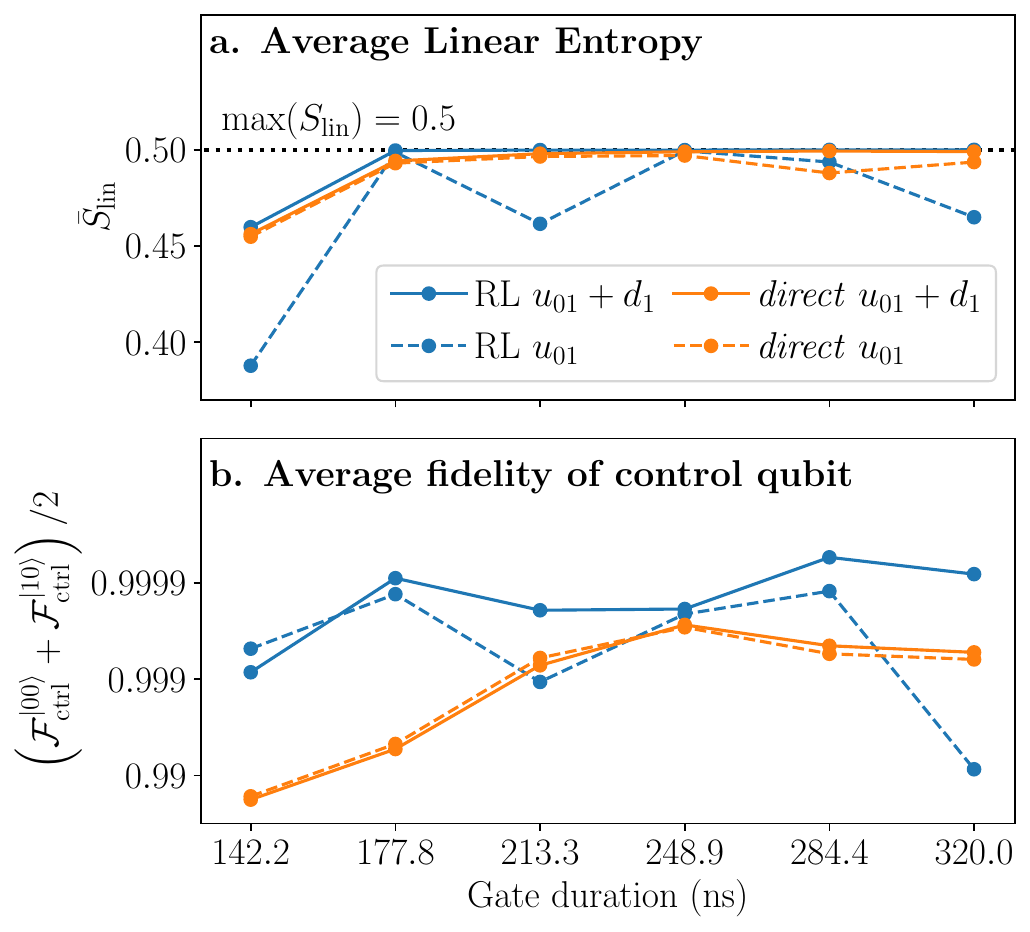}
    \caption{
    \textbf{Effect of removing target qubit drive $d_1$ from optimized $ZX(\pi/2)$ pulses.} 
    a) Average linear entropy $\bar{S}_{\rm lin}$ as defined in Sec.~\ref{subsec:entanglement}.
    b) Fidelity of control qubit averaged over initial states $\ket{00}$ and $\ket{10}$. The target qubit drive practically only affects the target qubit rotation in the \emph{direct} scheme. Meanwhile, the RL agent discovers solutions where this on-resonance drive works in tandem with the cross-resonance drive to generate entanglement and rotate the control qubit. 
    }
    \label{fig:roles_of_drives}
\end{figure}

We can further highlight the novelty in control solutions found by our RL agent by analyzing the role played by each drive in implementing a two-qubit operation. We do so by taking the each optimized pulse sequence, removing the on-resonance component $d_1$, and comparing the left-over cross-resonance component $u_{01}$ with the original pulse. In the weak drive regime, we expect the cross-resonance drive $u_{01}$ to be the sole entanglement generator and the on-resonance drive $d_1$ to only affect the local rotation of the target qubit. By examining the changes in the linear entropy and the qubit control fidelity when the on-resonance drive is removed, we confirm that, in the \emph{direct} scheme, $d_1$ has little to no effect on the entanglement generated and the motion of the control qubit, as illustrated in by the overlapping orange curves in Fig.~\ref{fig:roles_of_drives}. This suggests that optimizing the Gaussian Square pulses in the \emph{direct} scheme leads to control solutions exhibiting a clear separation of roles: the cross-resonance drive generates almost all entanglement and ensures that the control qubit ends up in the intended state; meanwhile, the on-resonance drive focuses on correcting the target qubit state. Interestingly, our RL agent discovers additional strategies where these roles get mixed up to different degrees, represented by how much the blue curves in Fig.~\ref{fig:roles_of_drives} deviate from one another. In these cases, the on-resonance drive $d_1$ actually works in tandem with the cross-resonance drive $u_{01}$ to generate entanglement and rotate the control qubit. Such a behavior can be attributed to high driving amplitudes which activate interactions beyond the desired $ZX$ term, leading to the  observed novelty in solutions discovered by our RL agent.

% Let us try to understand the role of each drive in implementing an entangling operation, by studying the effect of removing the on-resonance drive $d_1$ on the target qubit in the \emph{direct} and RL control solutions. Since this drive is expected to mainly affect the local rotation of the target qubit, we examine the changes in entanglement generated and the fidelity of the control qubit in its absence. As illustrated in Fig.~\ref{fig:roles_of_drives}, $d_1$ has little to no effect on the entanglement entropy and the control qubit fidelity in the \emph{direct} scheme (see orange curves). This suggests that optimizing the Gaussian Square pulses leads to solutions where there is a clear separation of roles: the cross-resonance drive generates almost all entanglement and ensures that the control qubit ends up in the intended state; meanwhile, the on-resonance drive focuses on correcting the target qubit state. In addition to this type of solution, our RL agent discovers strategies where these roles get mixed up to different degrees, represented by how much the blue curves in Fig.~\ref{fig:roles_of_drives} deviate from one another. In these cases, the target qubit drive works in tandem with the cross-resonance drive to generate entanglement and rotate the control qubit. Such a behavior is unexpected, from the perspective of the \textit{direct} scheme, and highlights the novelty in solutions discovered by our RL agent.

\subsection{Robustness of optimized pulses}\label{subsec:robustness}

\begin{figure}[t!]
    \centering
    \includegraphics[width=0.49\textwidth]{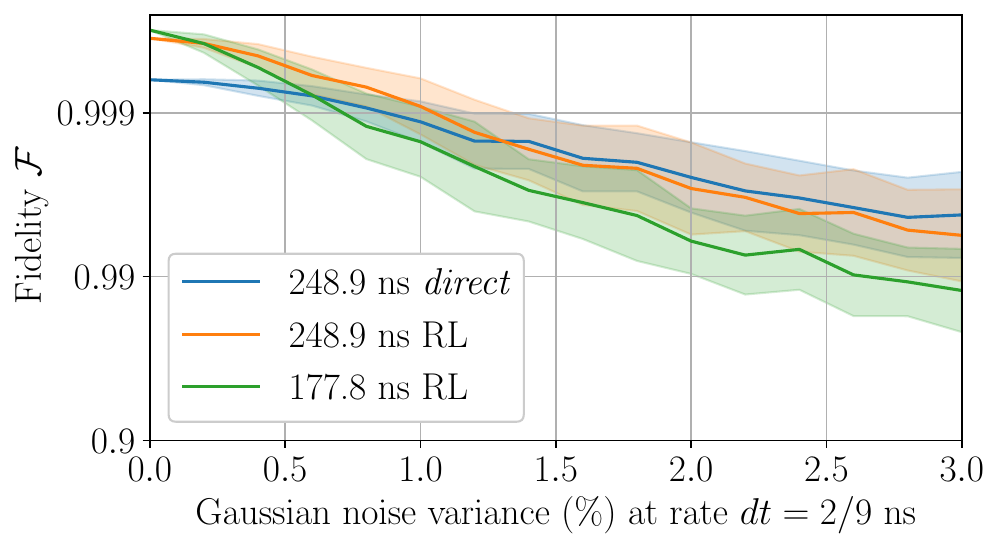}
    \caption{\textbf{Robustness of optimized $ZX(\pi/2)$ pulses to short-timescale stochastic noise.} 
    We simulate stochastic noise by adding uncertainty to system parameters at inverse sampling rate $dt=2/9$ ns during evaluation. RL-design pulses outperform the \emph{direct} scheme up to 0.5\% noise. For larger noise, the same-duration RL solution behaves similarly to its \emph{direct} counterpart whereas the shorter-duration RL solution degrades at a faster rate. RL training hyperparameters are given in the ``Fixed Environment" section of Table~\ref{tab:hyperparameters}.
    }
    \label{fig:robustness}
\end{figure}

When implemented on a real device, the performance of optimized control solutions inevitably suffers from a variety of error sources such as imperfect controls and noisy readouts. We simulate these effects by introducing Gaussian fluctuation on system parameters $\vec{p}$; this allows us to assess the robustness of the control solutions discussed so far. Specifically, at every step of the size of the inverse sampling rate $dt=2/9$ ns, fluctuations are sampled from a zero mean Gaussian distribution with a standard deviation of up to 3\% of the original system parameters $\vec{p}_0$ listed in Table~\ref{tab:deviceparam}. For each deviation value, we collect 50 samples and report the results for optimized $ZX(\pi/2)$ pulses in Fig.~\ref{fig:robustness}. RL-designed pulses outperform the \emph{direct} implementation up to 1\% deviation at the same gate duration (248.9 ns) and up to 0.5\% deviation at shorter gate duration (177.8 ns). The shorter gate duration RL solution degrades quickly as the deviation increases, possibly due to large jumps in its non-smooth PWC waveform. Since the 248.9 ns RL pulse has a much lower amplitude, the jumps in its PWC waveform are not as detrimental. As a result, it shows no discernible difference in performance degradation rate, when compared to its smooth Gaussian Square waveform counterpart in the \emph{direct} scheme. Overall, the mean fidelities drop to around 99-99.5\% at a 3\% deviation, which translates to about 66 kHZ for the coupling value and about several MHz for the remaining parameters. These results suggest a trade-off in the RL solutions between fidelity and gate duration in the presence of stochastic noise, which could stem from the non-smooth feature of PWC pulses as well as proximity to the smallest duration necessary to generate sufficient entanglement.

\subsection{Adapting to drifting system characteristics}\label{subsec:adaptetosystemdrift}

\begin{figure}[t!]
    \centering
    \includegraphics[width=0.48\textwidth]{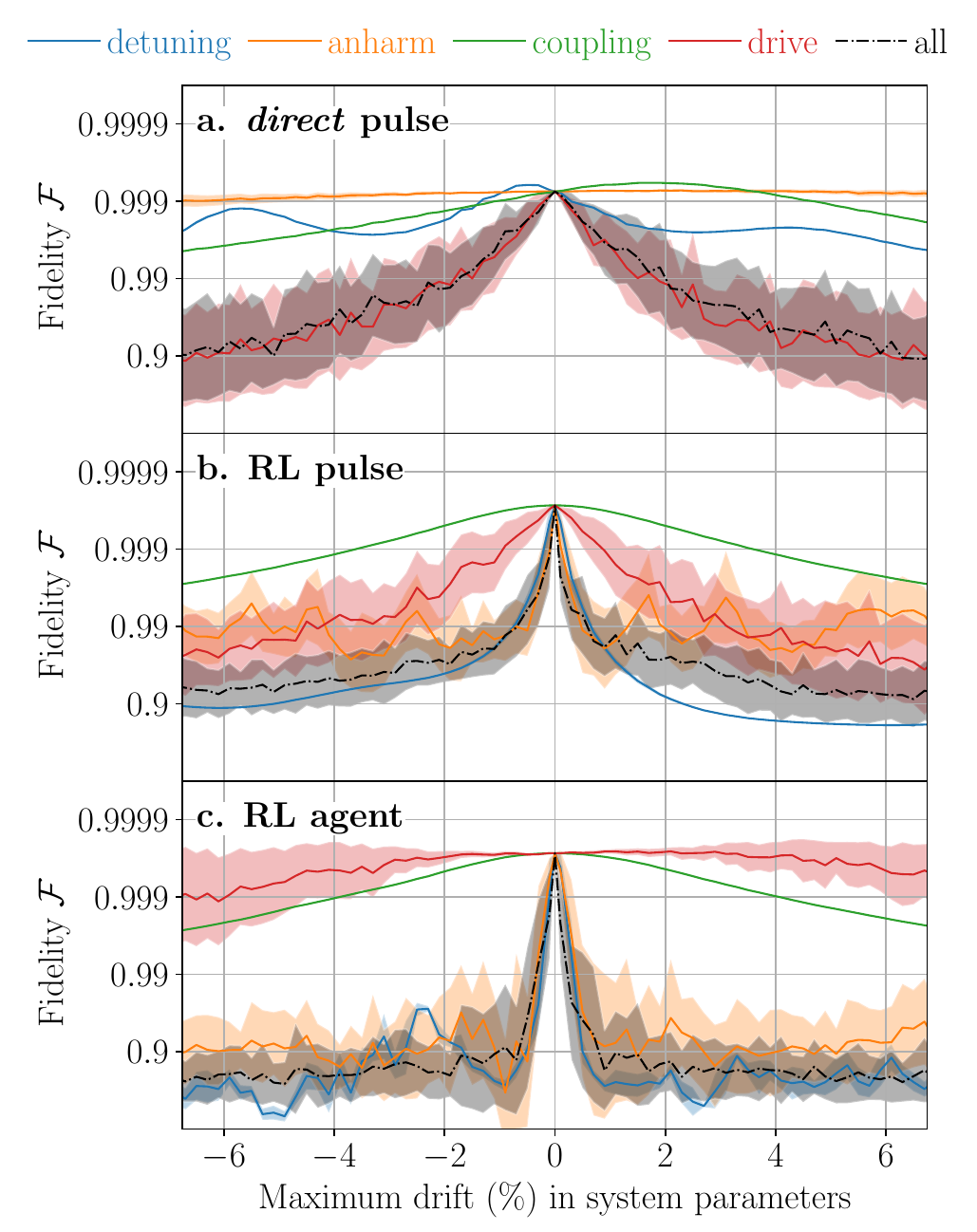}
    \caption{\textbf{Fidelity of fixed-environment control solutions in the presence of system drifts.} We sample drifts in system parameters (see legend) of a single type (solid curves) and of all types simultaneously (dashed black curve), and then bin the data points according to the maximum drift. The binned mean fidelities (curves) and their standard deviations (shaded areas) are displayed.
    (a-b) Fidelity of pulses from \emph{direct} implementation and RL optimization are evaluated on new environments. The \textit{direct} pulse is susceptible mostly only to drift in drive strength, whereas the RL solution is susceptible to drift in all parameters. 
    (c) Fidelity of adaptive pulse found by RL agent via interaction with each new environment; solution remains susceptible to drift in detuning and anharmonicity while generalizing well for drift in drive strength.
    Data represents optimizing a 248.9 ns CNOT pulse, with RL training hyperparameters given in the ``Drifting Environment" section of Table~\ref{tab:hyperparameters}.
    }
    \label{fig:generalization_fixedparam}
\end{figure}

In addition to the Markovian, short-timescale fluctuation, systems characteristics of superconducting transmon qubits are also known to drift over a longer time, owing to, e.g., dilution refrigerator temperature or fabrication defects. Therefore, in the following, we explore the idea of generalizing a single RL agent to a range of system parameters so that no additional training is required when re-calibration is needed \cite{preti2022continuous}. We additionally discover that knowledge of the change in system parameters can be utilized to strengthen the generalization capability of the agent. Finally, from this point on, we will switch to the objective of learning the CNOT gate to further highlight our approach's applicability to different target gates. Traditionally, a CNOT gate can be achieved by performing a $ZX(\pi/2)$ gate followed by a target qubit rotation. Since our RL agent learns to perform these gates from scratch, its control solution for the CNOT gate does not have to follow the aforementioned decomposition and can be learned directly, resulting in pulses reported in Fig.~\ref{fig:3drives_pulse}.

\subsubsection{Evaluating generalizability}
We first discuss our method of evaluating the generalizability of an RL agent on multiple systems whose parameters have drifted from the original values $\vec{p}_0$. We study two main situations, when only a single type of parameter has changed, i.e.,
\begin{eqnarray}
    \text{only coupling: }&& J\rightarrow J+\Delta J, \\
    \text{only drive strengths: }&& 
    \begin{cases}
        \Omega_{d_0} \rightarrow \Omega_{d_0} + \Delta \Omega_{d_0}, \\
        \Omega_{d_1} \rightarrow \Omega_{d_1} + \Delta \Omega_{d_1}, \\
        \Omega_{u_{01}}\rightarrow \Omega_{u_{01}}+\Delta \Omega_{u_{01}}, \\
        \Omega_{u_{10}}\rightarrow \Omega_{u_{10}}+\Delta \Omega_{u_{10}},
		 \end{cases} 
\end{eqnarray}
or when all parameters have changed. We gather the drifts in system parameters into a \textit{context} vector
\begin{eqnarray}
    \frac{\Delta\vec{p}}{\vec{p}_0} = \left[\frac{\Delta J}{J}, \frac{\Delta \Omega_{u_{01}}}{\Omega_{u_{01}}}, \dots \right].
\end{eqnarray}
We then sample these changes randomly, evaluate the fidelity of the tailored control solutions, and bin these data points according to the drift with largest absolute value. Using a bin-size of 0.2\%, we collect samples such that the number of data points increases from $\sim15$ points in the central bins to $\sim60$ points in the $\pm7\%$ bins. Finally, we study the binned mean and standard deviation of the fidelity as a function of the maximum drift.

We can now examine the generalizability of the control solutions discussed so far, namely the \emph{direct} and RL pulses optimized for the original system parameters, as shown in Fig.~\ref{fig:generalization_fixedparam}. We focus on a 6\% range of maximum drift on each side of the original values. In our simulation, since the coupling value is much smaller, the drift of the same percentages for this parameter has significantly less effect than the others (green curves). While generalizing well for most parameters due to its smoothness, the \emph{direct} pulse in Fig.~\ref{fig:generalization_fixedparam}a is highly susceptible to drift in drive strength as the rotation angle is directly affected. The RL-designed pulse in Fig.~\ref{fig:generalization_fixedparam}b, on the other hand, degrades quickly for drifts in all parameters as expected for a non-smooth PWC waveform. 

Adaptive control solutions from the RL agent for different system parameters behave more interestingly, which can be seen in Fig.~\ref{fig:generalization_fixedparam}c. We find a large susceptibility to changes in detuning and anharmonicity, likely because they directly change the spacing between transmon energy levels and fundamentally alter the system's internal physics, e.g., the resonant frequencies. Meanwhile, the drive strengths are connected to external controls which the RL agent has direct influence over, resulting in a much more robust behavior. Overall, previous control solutions exhibit poor generalization performance when drift in \textit{all} system parameters is considered.

\subsubsection{Learning to generalize}

\begin{figure}[t!]
    \centering
    \includegraphics[width=0.48\textwidth]{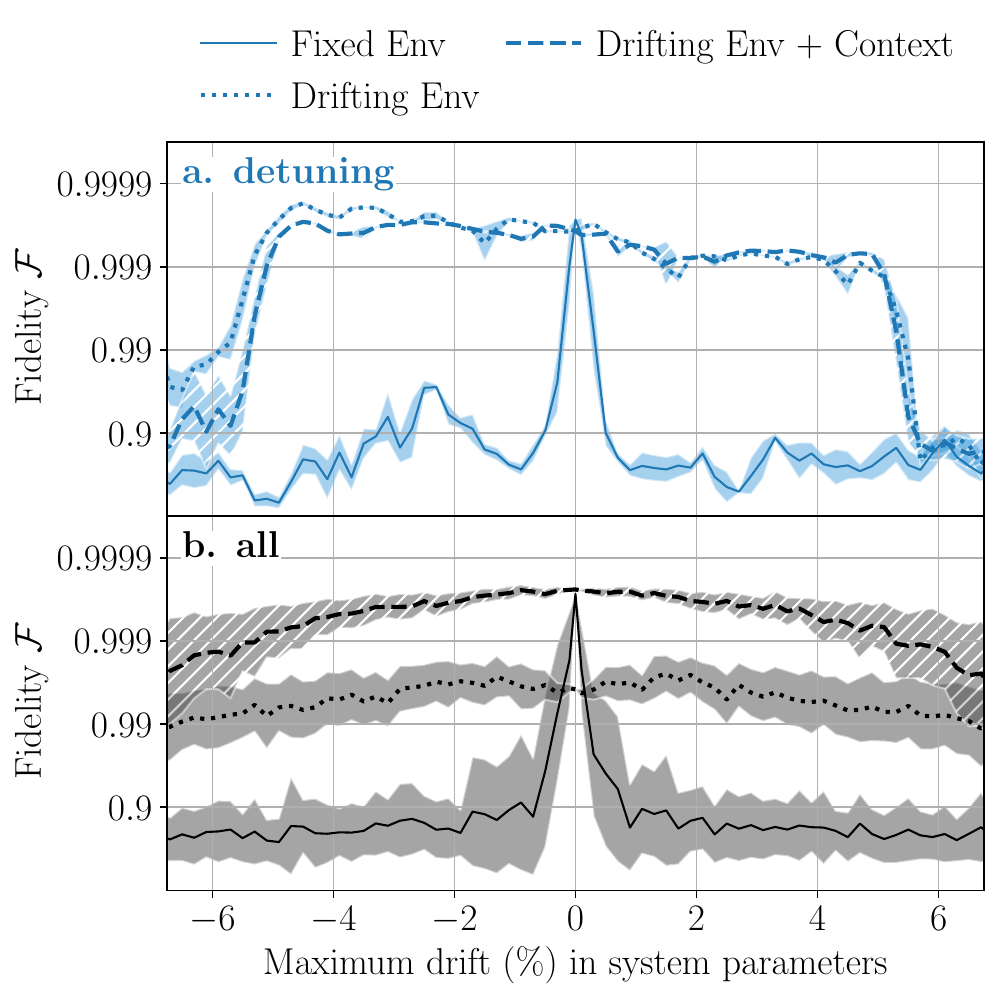}
    \caption{\textbf{Improved generalization fidelity when using augmented RL approach.} 
   As reference, we import fidelity curves (solid) from an agent trained on a fixed environment from Fig.~\ref{fig:generalization_fixedparam}c. We display improved results for training on an environment with drifting system parameters, when the agent has no knowledge (dotted) or full knowledge (dashed) of the drifting parameters, i.e., \textit{context} information. 
    (a) Drift in detuning frequency only. Training on environments with drifting system parameters is sufficient in improving the fidelity to $\gtrsim99.9\%$ within a 5\% drift. Having \textit{context} information provides a slight improvement in performance while cutting training time in half. 
    (b) Drift in all parameters. Context is essential to stabilize training, and provides the best generalization result. 
    RL training hyperparameters are given in the ``Drifting Environment" section of Table~\ref{tab:hyperparameters}.
    }
    \label{fig:generalization_capability}
\end{figure}

We now identify several ingredients necessary for improving the generalization performance of our agent. We start with a simpler problem where we task the agent to adapt to drift in detuning while all parameters remain fixed. By simply allowing the agent to interact with many systems with different detuning values during training, we observe an immediate and significant performance increase (cf.~dotted blue curve in Fig.~\ref{fig:generalization_capability}a). We achieve this by sampling the detuning from a uniform distribution of a 5\% range around its original value at the beginning of each episode. As a result, our agent can observe the changing effects of its action on drifted systems, and thus, learn to adapt accordingly. To further help the agent discern different systems more effectively, we can provide it with specific knowledge about the system with which it is currently interacting, namely the size of the detuning drift relative to its original value. We refer to this piece of information as \emph{context} and this additional input to the agent remains constant within each episode, implying that the system characteristics remain constant throughout the entire gate duration. While having little effect in generalization performance here (dashed blue curve), the training time needed for a \textit{context}-aware agent is actually cut in half. 

We apply our findings to the full problem with drift in all parameters and plot the result in Fig.~\ref{fig:generalization_capability}b. Training on a drifting environment remains necessary but is no longer sufficient to achieve good generalization results (dotted black curve). In fact, the best result we report when training without context is obtained in only a few training iterations; after that, the performance drops precipitously (cf.~black curve in Fig.~\ref{fig:learning_curves}b). We believe that feedback with the quantum state is no longer enough for our agent to distinguish different environments. For example, different environments can get to the same state with a different set of actions. Therefore, without additional information, the agent encounters a good amount of confusion during training. 

Providing our agent with \textit{context} information about the drifts in all parameters greatly alleviates the problem (dashed black curve). Here, instead of sampling from a uniform distribution as in the previous case, we sample drifts for all parameters from a zero-mean Gaussian distribution of a 2\% standard deviation, in order to highlight the effect on the generalization results. Indeed, instead of collapsing at 5\% drift (blue dashed curve), the fidelity in this case gradually decreases (black dashed curve) since the agent actually gets to interact with system drifts beyond 5\% during training under the Gaussian distribution. Overall, extending the training environment to include system drifts and providing the agent with \textit{context} information about those drifts significantly stabilizes the generalization task when all system parameters are involved. More importantly, the resultant agent can immediately propose pulses with 99.9\% fidelity at up to 4\% drift without any further training. These simulations justify the suggested practicality of RL in the presence of a reasonable system drift on near-term devices.

\begin{figure}[t!]
    \centering
    \includegraphics[width=0.48\textwidth]{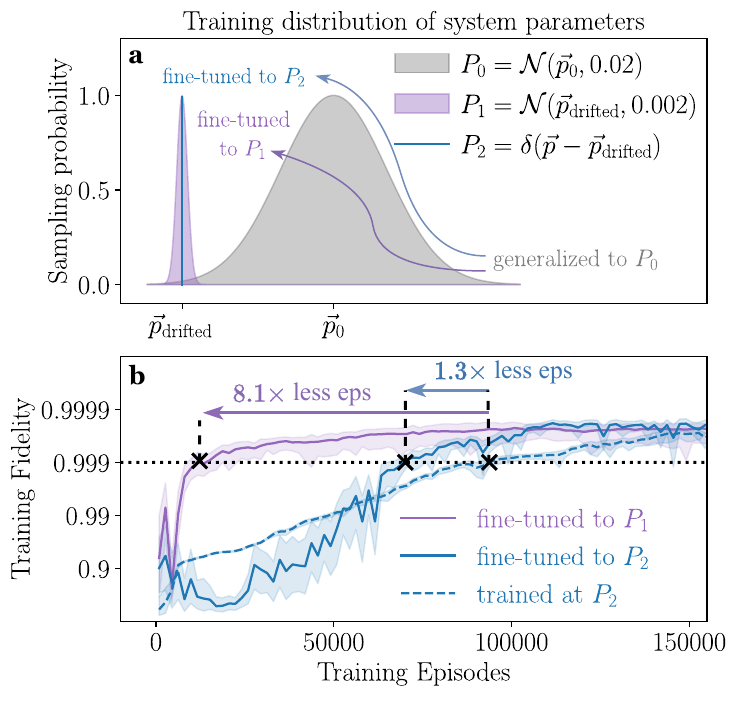}
    \caption{\textbf{Reduction in training episodes when using a generalized agent to fine-tune.} a) Training distribution of system parameters for three cases: generalizing to a Gaussian distribution $P_0$ of 2\% standard deviation (gray) as a starting point (same agent for dashed black curve in Fig.~\ref{fig:generalization_capability}b), then either fine-tuning to a narrower distribution $P_1$ of 0.2\% standard deviation (purple), or fine-tuning to a fixed environment, i.e., a delta distribution $P_2$ (blue). $P_0$ is centered at the original system parameters $\vec{p}_0$ whereas $P_1$ and $P_2$ are centered at $\vec{p}_{\rm drifted}$.
    b) Mean training fidelity as a function of training episodes for comparison between fine-tuning (to $P_1$  and $P_2$) and training from scratch (under $P_2$).  Shaded region is defined by the minimum and maximum training fidelity. For each case, we perform 3 runs with different seeds and report the best learning curve. RL training hyperparameters are given in the ``Drifting Environment" section of Table~\ref{tab:hyperparameters}. Fine-tuning to $P_1$ and $P_2$ offers a $8.1\times$ and $1.3\times$ reduction in episodes required to reach 99.9\% mean fidelity, implying great potential for transfer learning.
    }
    \label{fig:transfer_learning}
\end{figure}

When dealing with a more substantial drift, we find that fine-tuning is necessary to achieve 99.9\% fidelity, although the number of training episodes required can be notably less than when starting from scratch. To investigate this phenomenon, we first reiterate that the RL agent used in Fig.~\ref{fig:generalization_capability}b has been trained to generalize to system parameters sampled from a Gaussian distribution of 2\% standard deviation around the original values $\vec{p}_0$, as in Fig.~\ref{fig:transfer_learning}a (grey distribution). We then select a set of drifted system parameters, denoted as $\vec{p}_{\rm drifted}$ \footnote{Relative changes in system parameters in the following order $[\Omega_{d_0},\Omega_{u_{01}},\Omega_{d_1},\Omega_{u_{10}}]$, $[\delta_0,\delta_1]$, $[\alpha_0,\alpha_1]$, $J$, are $[-0.00722279, 0.02263578, -0.0283148, -0.02048592]$, $[\textbf{-0.05684035},  0.05331411]$, $[-0.04396466,  0.05445264]$, $0.04959622$, where the maximum drift is $\approx-5.7\%$.} with maximum drift of $-5.7\%$, for which the generalized agent suggests a control solution achieving 95\% fidelity. Subsequently, we specialize our agent by training it on a fixed environment with these specific system parameters (blue delta distribution). 

Comparing this approach to training a separate agent entirely from scratch within the same environment, we observe a reduction of $1.3\times$ in the number of training episodes required to reach 99.9\% fidelity, as illustrated in Fig.~\ref{fig:transfer_learning}b. Intriguingly, when we instead specialize our agent to a drifting environment characterized by a Gaussian distribution of 0.2\% standard deviation and mean at $\vec{p}_{\rm drifted}$ (cf.~purple distribution in Fig.~\ref{fig:transfer_learning}a), the episode reduction jumps to $8.1\times$. These preliminary findings hint at the potential for substantial transfer learning in certain cases. However, further investigations are needed to understand the underlying causes for such a wide range of effectiveness, which we plan to pursue in future studies.

\section{Conclusion and Outlook} \label{sec:discussion}

In this study, we have showcased the advantages of harnessing reinforcement learning for the design of cross-resonance gates, fully independent of known theoretical protocols and pre-existing error suppression techniques. Our unbiased approach employs an off-policy agent to customize continuous control parameters, shaping complex-valued pulses concurrently for both the cross-resonance and the target on-resonance drives. Compared to established optimal control methods, RL has the advantage of i) being suited for closed-loop optimization due to its model-free nature; ii) it is capable of non-local exploration; and iii) it generates a representation of gained knowledge as a valuable by-product.

Using the RL methodology, we demonstrated the discovery of novel control solutions that fundamentally differ from conventional error suppression techniques for two-qubit gates, such as \textit{direct} and \textit{echoed}  schemes, while surpassing them in both fidelity and execution time. At the typical gate duration of 248.9 ns (for transmon devices) where the \emph{direct} and \textit{echoed} schemes achieve $99.937\%$ and $99.501\%$ fidelity respectively, RL-designed solutions can cut the error in about half compared to the better scheme, achieving $\mathcal{F}_{\text{RL}}\gtrsim99.966\%$, for both $ZX(\pi/2)$ and $\text{CNOT}$ gates. Not only that, our agent identified a potential maximum reduction of 30\% in gate duration, while maintaining the same level of fidelity exceeding $99.9\%$. This can be attributed to the flexibility of the piece-wise constant ansatz capable of managing leakage out of the computational subspace, as well as unwanted coherent processes that inevitably arise at large drive amplitudes. 

Furthermore, we illustrated the possibility of augmenting our approach to enable our agent to flexibly adapt its design capability to accommodate drifts in the underlying hardware. We found that exposing the agent to an environment with drifting system parameters during training while providing it with \textit{context} information about these drifts, allows our agent to learn the appropriate control solutions and generalize well across a range of drifted system parameters. Concretely, our \textit{context-}aware agent can readily propose control solutions with $\sim 99.9\%$ fidelity when all system parameters, including detuning, anharmonicity, coupling strength, and drive strength, are allowed to drift within a 4\% range around their original values. In instances of more substantial drifts, our generalized agent serves as a valuable starting point for fine-tuning, resulting in a remarkable $1.3-8.1\times$ acceleration in optimization iterations when compared to starting from scratch.

Based on these findings, we can assert that the RL approach alleviates the necessity for a precise model, presenting a versatile framework applicable to designing various cross-resonance-based gates. When combined with piece-wise constant protocol space, our RL agent demonstrates its capacity to devise innovative pulse shapes that surpass the capabilities of conventional ans\"atze in terms of both fidelity and gate execution duration. The quest for shorter, high-fidelity pulses is particularly significant, given that various calibration methods are nearing the coherence limit imposed by state-of-the-art gate duration and qubit relaxation times. Furthermore, our \textit{context}-aware RL approach effectively addresses hardware drifts, indicating the possibility of reducing and even eliminating additional training, and thus expensive calibration experiments, as long as system characteristics remain within a reasonable range.

When applied to experiments conducted on real-world hardware, our off-policy method carries the potential for significant data efficiency gains, as the agent can be trained on data collected by any policy. Consequently, while the initial training phase may incur high costs, subsequent retraining can be expedited thanks to the collected dataset. Additionally, actual drives delivered to the qubits are generally smoothed out from the raw jagged PWC input pulses, which should enhance the robustness of the optimized solutions to control fluctuations. 

As the quantum computing community progresses toward larger platforms, the capacity of a single agent to extend its design capabilities across diverse system characteristics becomes increasingly pivotal for scalability \cite{preti2022continuous}. In fact, as the number of qubits grows, it becomes inevitable that certain qubits will exhibit overlapping system characteristics \cite{berke2022transmon}. In such a scenario, our \textit{context}-aware agent, trained to generalize within a specific region of system parameters, can readily be applied to a group of qubits sharing similar characteristics. Moreover, these experiments can be conducted simultaneously, as qubits with akin parameters are likely to be positioned at a considerable distance from each other in the first place, further enhancing the efficiency of our RL agent.

In the immediate future, we are eager to integrate our approach into established gate optimization procedures for superconducting devices, as well as extending its utility to various quantum computing platforms. At the same time, we also aim to broaden the applicability of our RL agent to handle more intricate operations, such as the SWAP gate, multi-qubit gates, or gates on qudits. With the recent advancement enabling better control of the $\ket{1}\leftrightarrow\ket{2}$ transition~\cite{qutrit_Goss_2022,twoqutritgate_mahadevanlupascu_2023} and the potential of improving quantum speed limit by expanding beyond the qubit subspace~\cite{basyildiz2023speed}, the synthesis of even faster qubit gates as well as qutrit gates emerges as an intriguing and imminent application for our RL protocol. On the algorithmic front, we emphasize the significance of enhancing generic RL algorithms through generalization and transfer learning techniques to bolster the method's practicability, especially for large-scale platforms. With the field of reinforcement learning, and machine learning in general, growing at an unprecedented rate, we hope to continue leveraging these powerful advancements toward the development of practical quantum computers.

\section*{Acknowledgements}

This material is based upon work supported by the U.S. Department of Energy, Office of Science, National Quantum Information Science Research Centers, Quantum Systems Accelerator (HNN, KBW). Additional support is acknowledged from the Helmholtz Initiative and Networking Fund, grant no.~VH-NG-1711 (MS), the European Union (ERC, QuSimCtrl, 101113633) (MB). Views and opinions expressed are however those of the authors only and do not necessarily reflect those of the European Union or the European Research Council Executive Agency; neither the European Union nor the granting authority can be held responsible for them.

The authors gratefully acknowledge the Gauss Centre for Supercomputing e.V. (www.gauss-centre.eu) for funding this project by providing computing time through the John von Neumann Institute for Computing (NIC) on the GCS Supercomputer JUWELS \cite{JUWELS} at J\"ulich Supercomputing Centre (JSC).

Finally, the authors thank Haoran Liao and John Paul Marceaux for helpful conversation on various subjects pertained to this work.

\appendix 

\begin{table*}[t]
\begin{centering}
\begin{tabular}{|c||c|c|c||c|c||c|c|} 
 \multicolumn{1}{c}{} &  \multicolumn{3}{c}{\textbf{Fixed Environment}} &\multicolumn{2}{c}{\textbf{Drifting Environment}} &\multicolumn{2}{c}{\textbf{3 drives}}\\
\hline \textbf{Environment} &\multicolumn{3}{c||}{} &\multicolumn{2}{c||}{} &\multicolumn{2}{c|}{}\\
\hline Target gate & $IX(\pi/2)$ &\multicolumn{2}{c||}{$ZX(\pi/2)$,CNOT} &\multicolumn{2}{c||}{CNOT} &\multicolumn{2}{c|}{$ZX(\pi/2)$,CNOT} \\
\hline Number of segments& 9& 20&20&20& 28* &\multicolumn{2}{c|}{20} \\ 
\hline Drifting system parameters& \multicolumn{3}{c||}{-} & detuning &  all &\multicolumn{2}{c|}{-} \\
\hline  Sampling distribution & \multicolumn{3}{c||}{-} &uniform&normal &\multicolumn{2}{c|}{-} \\ 
\hline Sampling variance&  \multicolumn{3}{c||}{-} &5\%&2\% &\multicolumn{2}{c|}{-} \\
\hline &  & 320.0 ns& 213.3 ns &\multicolumn{2}{c||}{} &&\\
 Duration & 10 ns & 284.4 ns& 177.7 ns & \multicolumn{2}{c||}{248.9 ns}& 248.9 ns & 177.8 ns \\
 & & 248.9 ns& 142.2 ns & \multicolumn{2}{c||}{} &&\\
\hline Drives & $d_1$ & \multicolumn{2}{c||}{$u_{01},d_1$} &\multicolumn{2}{c||}{$u_{01},d_1$} &\multicolumn{2}{c|}{$d_0,u_{01},d_1$}\\
\hline Action window $w_u$& 0.4& 0.1&0.2&\multicolumn{2}{c||}{0.1} &0.1&0.15\\
\hline Action window $w_d$& 0.13& 0.01&0.02&\multicolumn{2}{c||}{0.01} &0.01&0.015 \\
\hline \hline \textbf{DDPG} &   \multicolumn{3}{c||}{} &\multicolumn{2}{c||}{} &\multicolumn{2}{c|}{}\\
\hline Hidden layers & {100,200,100}  &\multicolumn{2}{c||}{800,400,200} &\multicolumn{2}{c||}{800,800,800} &\multicolumn{2}{c|}{800,400,200} \\
\hline Activation & \multicolumn{7}{c|}{relu}  \\
\hline Learning rate & \multicolumn{7}{c|}{1e-4}  \\
\hline Train batch size & \multicolumn{7}{c|}{64} \\
\hline Soft update parameter $\kappa$ & \multicolumn{7}{c|}{0.002} \\
\hline Buffer capacity & \multicolumn{7}{c|}{100000} \\
\hline Initial buffer & \multicolumn{7}{c|}{10000} \\
\hline 
\end{tabular}
\par\end{centering}
\caption{Training hyperparameters for training an RL agent to design quantum gates on simulated transmon environment. Unmentioned hyperparameters necessary for the DDPG algorithm are set to their default values in RLlib's implementation~\cite{liang2018rllib}, version 2.0.0. 
*When studying the adaptability of our RL agent to drifting system characteristics, we discover a particular region around +2\% drift on all system parameters where the 20-segment ansatz yields no solution with fidelity better than 99.5\%. The problem goes away when we increase the number of segments to 28, whose result was reported in the main text.} 
\label{tab:hyperparameters}
\end{table*}

\section{Simulated environment details}

\begin{table}[t!]
\begin{center}
    \begin{tabular}{|c|c|c|} 
        \hline & Symbols & Values (MHz) \\
        \hline \hline
        Drive strengths &  $\Omega_{u_{01}},\Omega_{d_0}$& 204.7  \\ 
        \hline & $\Omega_{u_{10}},\Omega_{d_1}$& 158.5 \\ 
        \hline \hline Detuning & $\delta_{0}$ & -86.6  \\
        \hline & $\delta_{1}$ & 0  \\
        \hline \hline Anharmonicity & $\alpha_{0}$ & -310.5  \\
        \hline & $\alpha_{1}$ & -313.9  \\
        \hline \hline Coupling & $J$ & 2.2  \\
        \hline
    \end{tabular}
    \caption{IBMQ Valencia system parameters extracted from Qiskit interface,  and used in this work. We denote the corresponding system parameter vector to be $\vec{p}_0=[J,\Omega_{u_{01}},\dots]$. Typical bare qubit frequencies have been translated into detuning values used in our simulation.}     \label{tab:deviceparam}
\end{center}
\end{table}

\subsection{Rotating frame transformation}\label{app:transformation}
We consider the single-transmon Hamiltonian in the lab frame $H_1^{\rm lab} = H_{\rm 1,sys}^{\rm lab} + H_{\rm 1,ctrl}^{\rm lab}(t)$, given in Sec.~\ref{subsec:singlequbitgates}. Under the rotating frame transformation $R=R(t)=e^{-i\omega_d t b^\dagger b}$, the ladder operators are mapped as follows
\begin{eqnarray}
    b^\dagger &\rightarrow& R^\dagger(t) b^\dagger R(t) = b^\dagger e^{i\omega_d t} \nonumber \\
    b &\rightarrow& R^\dagger(t) b R(t) = b e^{-i\omega_d t},
\end{eqnarray}
leaving the number-conserving terms, i.e., with equal numbers of $b^\dagger$ and $b$, unchanged. The derivative $-R^\dagger i\partial_t R = -\omega_d b^\dagger b$ contributes to the quadratic term, leading to the detuning $\delta=\omega-\omega_d$ when combined with the transformed system Hamiltonian:
\begin{eqnarray}
    R^{\dagger}(H_{\rm 1,sys}^{\rm lab}-i\partial_t)R &=&  \delta b^{\dagger}b+\frac{\alpha}{2}b^{\dagger}b^{\dagger}bb.
\end{eqnarray}
The control part then transforms as follows
\begin{eqnarray}
   R^{\dagger}H_{\rm 1,ctrl}^{\rm lab}(t)R &=& \Omega_d\Re\left(d(t)e^{i\omega_d t}\right)(b^{\dagger}e^{i\omega_d t} + be^{-i\omega_d t}) \nonumber \\
    &=& \frac{\Omega_d}{2} \left[ d(t)b^\dagger e^{2i\omega_d t} + d(t)b + {\rm h.c.}\right] \nonumber \\
    &\approx& \frac{\Omega_d}{2} \left[ d(t)b + {\rm h.c.}\right]
\end{eqnarray}
where in the last line, we have dropped terms rotating at twice the driving frequency $\omega_d$, which is commonly known as the Rotating Wave Approximation (RWA). Altogether, we arrive at the rotating frame Hamiltonian for single transmon $H_1(t)$ as in Eq.~\ref{eq:H1}. For two-transmon, the above calculation carries directly over.

\subsection{Virtual Z gates and angles} \label{app:virtualZ}

We first consider applying a virtual $Z$ gate only on the first qubit by augmenting the overlap $M=U_{\rm qubit}U^\dagger_{\rm target}$ as follows
\begin{eqnarray}
    M \rightarrow M'=V_Z(\theta_0)M = \left[\text{diag}(1,e^{i\theta_0})\otimes I\right] M.
\end{eqnarray}
While the first term in~Eq.~\ref{eq:Favg} remains unchanged: $\Tr(M'M'^\dagger)= \Tr(V_Z(\theta_0)MM^\dagger V_Z(-\theta_0)) $$= \Tr(MM^\dagger)$, the second term becomes
\begin{eqnarray}
    |\Tr(M')|^2 &=& \left|M_{00}+M_{11}+(M_{22}+M_{33})e^{i\theta_0} \right|^2 \nonumber \\
    &=& |M_{00}+M_{11}|^2 + |M_{22}+M_{33}|^2 \nonumber \\
    &+& (M_{00}+M_{11})^*(M_{22}+M_{33})e^{i\theta_0} \nonumber \\
    &+& (M_{00}+M_{11})(M_{22}+M_{33})^*e^{-i\theta_0} \nonumber \\
    &=& \text{const} + \mathcal{M}e^{i\theta_0} + \mathcal{M}^*e^{-i\theta_0},
\end{eqnarray}
where $M_{ij}$ are the matrix elements of $M$, and $\mathcal{M}=(M_{00}+M_{11})^*(M_{22}+M_{33})$. We find the optimal value of $\theta_0$ that maximizes the average gate fidelity by solving
\begin{eqnarray}
    \frac{\partial \mathcal{F}_{\rm avg}}{\partial \theta_0} = 0 
    \Rightarrow e^{2i\theta_0}=\frac{\mathcal{M}^*}{\mathcal{M}} \nonumber \\ \Rightarrow \tan \theta_0  = -\frac{\Im\mathcal{M}}{\Re\mathcal{M}}, \label{eq:virtualZangle}
\end{eqnarray}
where there exist two solutions within the $[-\pi,\pi]$ range. To ensure that we obtain the angle that maximizes the fidelity instead of minimizing it, we check that the solution angle of the above relation satisfies
\begin{eqnarray}
    \frac{\partial^2 \mathcal{F}_{\rm avg}}{\partial^2 \theta_0} &<& 0 \nonumber\\ 
    \Rightarrow \sin \theta_0 \Im \mathcal{M} &<& \cos \theta_0 \Re \mathcal{M}.
\end{eqnarray}
If this is not the case, we simply pick the other solution $\theta_0 - \text{sign}(\theta_0)\pi$. In the case of a virtual $Z$ gate only on the second qubit $V_Z(\theta_1) = \left[I\otimes \text{diag}(1,e^{i\theta_1})\right]$, we follow the same derivation with $\mathcal{M}=(M_{00}+M_{22})^*(M_{11}+M_{33})$. 

Obtaining the analytical optimal values for both angles simultaneously by solving 
\begin{eqnarray}
    (\theta^{\rm opt}_0,\theta^{\rm opt}_1) = \max_{\theta_0,\theta_1} \mathcal{F}_{\rm avg}(V_Z(\theta_1)V_Z(\theta_0)M),
\end{eqnarray}
is much more involved since the dependence of $\mathcal{F}_{\rm avg}$ on $(\theta_0,\theta_1)$ is significantly more complicated. However, we find that optimizing one angle at a time using Eq.~\ref{eq:virtualZangle} by solving
\begin{eqnarray}
    \theta^{\rm near-opt}_0 &=& \max_{\theta_0} \mathcal{F}_{\rm avg}(V_Z(\theta_1)M) \nonumber \\
    \text{then } \theta^{\rm near-opt}_1 &=& \max_{\theta_1} \mathcal{F}_{\rm avg}(V_Z(\theta_1)V_Z(\theta^*_0)M) \label{eq:virtualZprotocol}
\end{eqnarray}
yields a near-optimal solution $\bm{\theta}^* = (\theta^*_0,\theta^*_1)$ that is sufficiently accurate for the fidelity to reach the threshold we set: indeed, we find that the error in computing the maximum average fidelity using this protocol, as compared to numerically optimizing for both angles simultaneously, remains below $10^{-5}$, which is negligible for the fidelity levels discussed in our work.

In our simulation where we work in the rotating frame of the second transmon, more $Z$ error accumulates on the first transmon. We observe that optimizing for the larger angle yields a more accurate result, whence the order in Eq.~\ref{eq:virtualZprotocol}.

\subsection{Evolving method}

We provide details into our simulator of transmon quantum dynamics, which is unitary in the absence of decoherence processes. Under piece-wise-constant controls, the unitary map naturally simplifies into a product of time-local propagators 
\begin{equation}
    U = \prod_{n=1}^N U_n = U\left[N\Delta t,(N-1)\Delta t\right]\cdots U\left[\Delta t,0\right],
\end{equation}
where $\Delta t$ is the discretized time step, and $N$ is the number of segments in the PWC pulse. Each propagator, corresponding to each segment, can be computed by solving the time-dependent Schrödinger equation (TDSE)
\begin{equation}
    \frac{\partial U(t)}{\partial t} = -iH(t)U(t)
\end{equation}
from $t$ to $t+\Delta t$. For example, in the two-transmon setting, we set $H(t)=H_2(t)$ given  in Eq.~\ref{eq:H2total}. In general, the unitary is given by the following time-ordered integral 
\begin{equation}
    U_{\text{TDSE}}[t+\Delta t, t] =\mathcal{T} \exp\left[ -i\int_{t}^{t+\Delta t}H(t)dt \right],
\end{equation}
which we numerically obtain using QuTiP's TDSE solver~\cite{qutip_JOHANSSON20131234}. This is necessary because of the detuning phase factors $e^{i\delta t}$, even though the control fields $u$ and $d$ are constant within each segment. For the majority of this work, however, we focus on shaping only two control fields, $u_{01}$ and $d_1$, and the phase factors drop out in the frame rotating at the target qubit frequency. As the result, the piece-wise Hamiltonian is now constant and can be directly exponentiated to obtain the unitary that solves the time-independent Schrödinger equation (TISE)
\begin{equation}
    U_{\text{TISE}}[t+\Delta t, t] = \exp\left[ -iH(t)\Delta t \right],
\end{equation}
providing a significant computational speedup. We verified that TISE solution converges to TDSE solution as the step size approaches zero.

\begin{algorithm}[H]
\caption{Deep Deterministic Policy Gradient}
\label{alg:ddpg}
\begin{algorithmic}[1]
\Require Initial Q-network and policy parameters $\phi$ and $\theta$
\Require Initial target networks parameters $\phi'$ and $\theta'$
\For{step $= 1,2, \dots, M$}
    \State Reset environment to state $s_0$ \textbf{if} end of episode
    \State Sample an exploration noise $\mathcal{N}$
    \State Select action $a_i = \mu_\theta(s_i) + \mathcal{N}$
    \State Evolve the environment and observe $(r_i,s_{i+1})$
    \State Add tuple $(s_i, a_i, s_{i+1}, r_{i+1})$ to replay buffer $\mathcal{B}$
    \If{$\mathcal{B}$ has more than $N_{\rm initial}$ samples}
        \State Sample a batch $B$ of $(s_i, a_i, s_{i+1}, r_{i+1})$ from $\mathcal{B}$
        \State Compute targets $y_i = r_{i+1} + \gamma Q_{\phi'}(s_{i+1},\mu_{\theta'}(s_i))$
    \State Update Q-network by minimizing the loss:
    \begin{equation*}
        L_\phi=\frac{1}{|\mathcal{B}|}\sum_{i=1}^{|\mathcal{B}|}\left[y_i-Q_\phi(s_i, a_i)\right]^2
    \end{equation*}
    \State Update policy by maximizing the Q-values:
    \begin{equation*}
        L_\theta = - \frac{1}{|\mathcal{B}|} \sum_{i=1}^{|\mathcal{B}|} Q_\phi(s_i,\mu_\theta(s_i))     
    \end{equation*}
    \State Update target networks: 
    \begin{eqnarray}
        \theta' &\leftarrow& \kappa \theta + (1 - \kappa) \theta' \nonumber \\
        \phi' &\leftarrow& \kappa \phi + (1 - \kappa) \phi' \label{eq:polyak}
    \end{eqnarray}
\EndIf
\EndFor
\end{algorithmic}
\end{algorithm}

\section{DDPG Algorithm}\label{app:ddpg}

In this work, we employ Deep Deterministic Policy Gradient (DDPG), an off-policy Q-learning algorithm suitable for a continuous action space. We summarize the training procedure in Algorithm~\ref{alg:ddpg}.

The two neural networks for estimating the optimal Q-value and the agent's deterministic policy are randomly initialized at the beginning of training. For continuous action space, exploration is implemented by adding some noise $\mathcal{N}$ directly to the  policy network's output: $a_i = \mu_\theta (s_i) + \mathcal{N}$. The exploration noise is scaled down over time using the Ornstein-Uhlenbeck process as implemented in the DDPG paper~\cite{lillicrap2019continuous}.

Transitions $(s_i,a_i,s_{i+1},r_{i+1})$ are collected by following the agent's noisy policy and stored in a large replay buffer $\mathcal{B}$. For the first 10000 steps, the buffer is being filled without learning. After that, a batch of transitions are used along with two independent Adam optimizers to update both networks. To maintain quasi-stable targets throughout training, we soft-update both target networks $(\phi',\theta')$ via Polyak averaging, see Eq.~\ref{eq:polyak}.

We employ the DDPG (and TD3) implementation from RLlib, an open-source industry-grade library for RL~\cite{liang2018rllib}. We conduct a routine exploration of hyperparameters to identify an effective setting, which we maintaine consistently throughout the study. Due to the complex interplay between hyperparameters in high-dimensional analysis, altering one may impact others, making it challenging to provide a comprehensive account. Our focus is on identifying an effective set of hyperparameters and after that, minimizing additional adjustments to maintain the stability in our approach. The detailed hyperparameters used in this work are summarized in Table \ref{tab:hyperparameters}.  Any other hyperparameters not mentioned are unchanged from the default setting of RLlib version 2.0.0.

\section{Training procedure}\label{app:trainingprocedure}
\begin{figure}[t!]
    \centering
    \includegraphics[width=0.49\textwidth]{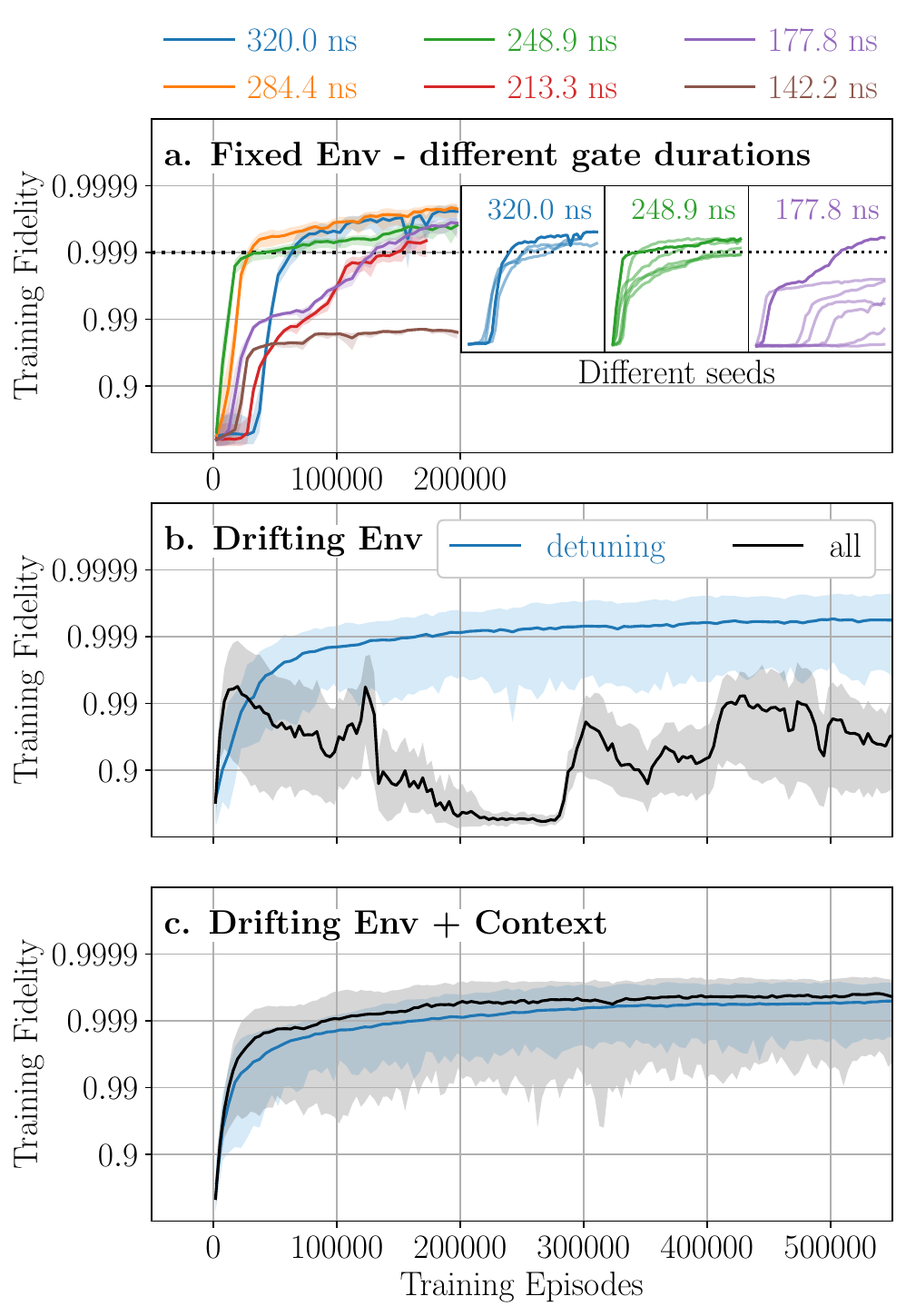}
    \caption{\textbf{Learning curves for main text results.} We display a moving average computed using a window of $\sim3500$ episodes and the shaded region is defined by the minimum and maximum training fidelity. a) Agents trained on a fixed environment for different gate durations consistently achieve $99.9\%$ fidelity for gate duration $\geq248.9$ ns after about 150,000 episodes. \textit{Inset:} Running each case for multiple seeds shows that the shorter the gate duration, the harder it is for the RL agent to learn. b-c) Agents trained on an environment with changing system parameters with and without context information provided. When only the detuning varies, context information helps accelerate learning although it not necessary for convergence (blue curves). When all parameters vary, \textit{context} information is key to stable training and convergence. Training converges around 500,000 episodes. All RL training hyperparameters are given in the Table~\ref{tab:hyperparameters}.
    }
    \label{fig:learning_curves}
\end{figure}

We report learning curves for the RL results discussed in the main text, plotting the mean fidelity of pulses encountered as a function of training episodes. Fig.~\ref{fig:learning_curves}a shows the learning curves for RL training on a fixed environment for different gate durations.  Our DDPG agent consistently finds $\geq99.9\%$ fidelity control solutions after about 150,000 episodes for gate durations $\geq248.9$ ns, which corresponds to about $18$ hours of training. Below this number, training becomes more challenging as we increase the action windows $w_u$ and $w_d$ to compensate for shorter physical time. An increase in action space often leads to exploding $Q$-values, resulting in a lower success rate over multiple runs. It should be pointed out that implementing the TD3 tricks~\cite{TD3_fujimoto2018addressing} stabilizes training but degrades the achievable fidelity slightly when compared to DDPG in general. 

Fig.~\ref{fig:learning_curves}b-c show learning curves for RL training on an environment with changing system parameters. Unlike both stable blue curves, the black curve is only stable when \textit{context} is included, suggesting the importance of \textit{context} information for adapting RL to a more realistic situation where \emph{all} system parameters can drift away from their original values. The generalized agent typically converges around 500,000 episodes, which corresponds to about 4 days of training.

Each training instance initiates 4 workers for sampling interaction with the simulated environment and 1 worker for agent training, utilizing 5 cores simultaneously. Without any significant difference in runtime, training is done either on a typical laptop (M1 3.2GHz or Intel i7 2.7GHz) or on a node within the JUWELS cluster (Intel Xeon Platinum 8168 2.7 GHz).

\section{Dynamics of optimized pulses}
\subsection{Leakage for RL pulses}
\begin{figure}[t!]
    \centering
    \includegraphics[width=0.48\textwidth]{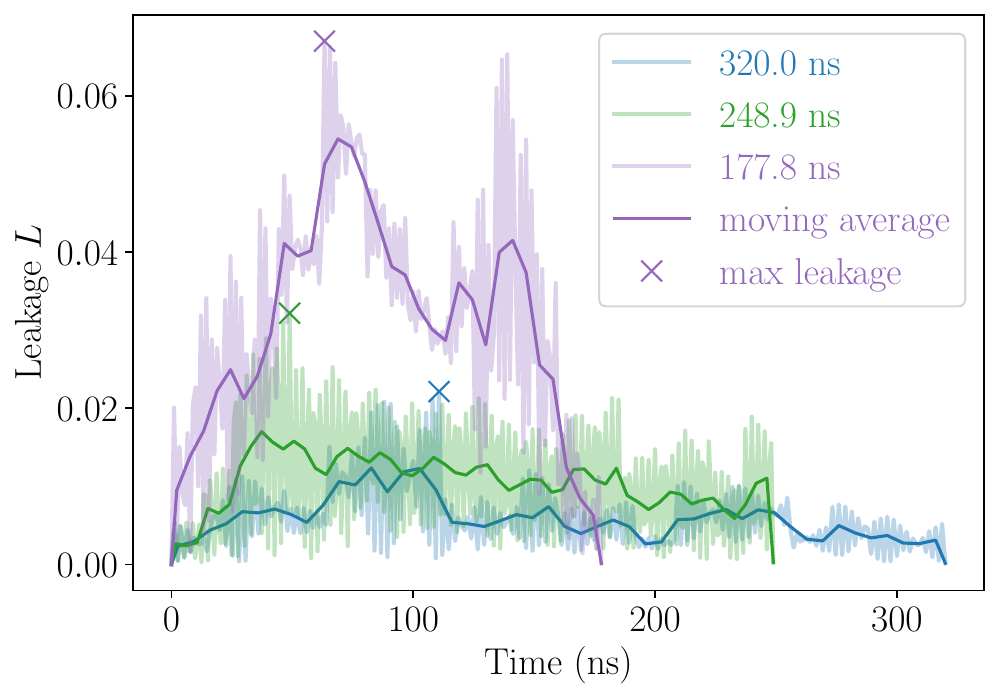}
    \caption{\textbf{Population leakage throughout gate duration for RL $ZX(\pi/2)$ pulses with different gate durations.} We display high-resolution evolution of the leakage value as well as its moving average. Crosses mark the maximum leakage data points reported in Fig.~\ref{fig:shortergatetime}. Larger population leakage is observed for shorter gate duration, which can be attributed to corresponding higher drive amplitude.
    }
    \label{fig:leakage_gatetimes}
\end{figure}
\begin{figure}[t!]
    \includegraphics[width=0.48\textwidth]{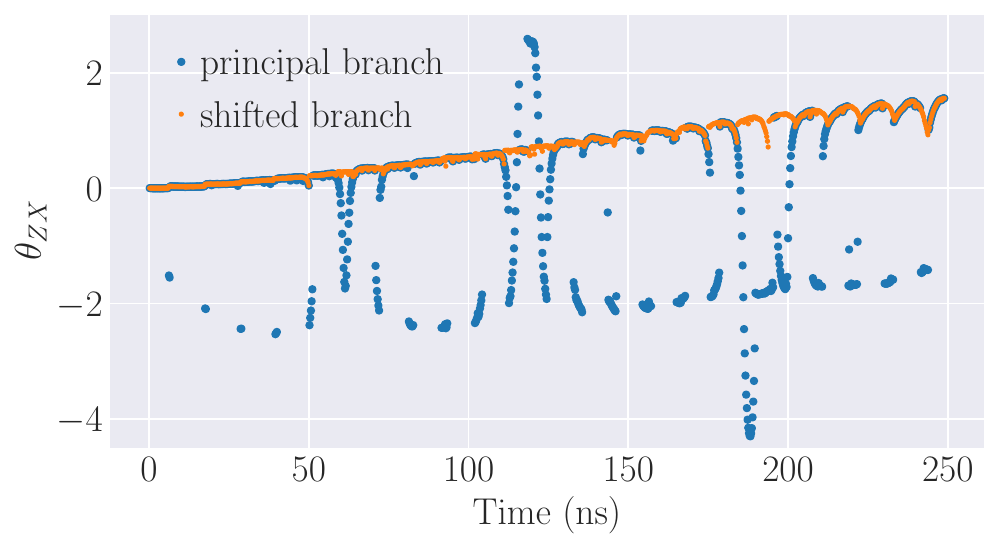}
    \caption{\textbf{Evolution of $ZX$ rotation angle under different branch cuts.} $\theta_{ZX}$ is first computed in the \textit{principal} branch cut (blue), i.e., without any phase shifts. Then, a phase shift is determined and added at every time step, resulting in the shifted branch cut (orange), which ensures a well-behaved evolution of rotation angles. Without shifting the branch cut, the observed large jumps obscure meaningful interpretation of the accumulated rotation angles. 
    }
    \label{fig:branchcuts}
\end{figure}

We report the full evolution of population leakage throughout gate duration for RL pulses with different gate durations in Fig.~\ref{fig:leakage_gatetimes}. As expected, shorter RL pulses exhibit higher leakage out of the computation qubit subspace in the intermediate time steps. Nevertheless, our RL agent manages to suppress almost all leakage at gate conclusion. We additionally observe fast oscillation in the leakage value, which is possibly due to off-resonant precession of the control qubit. Together, these observations provide a more detailed picture of the leakage process in the novel control solutions discovered by RL.

\subsection{Rotation angles}\label{app:rotation_angles}

\begin{figure*}[t!]
    \centerline{
        \includegraphics[width=1\textwidth]{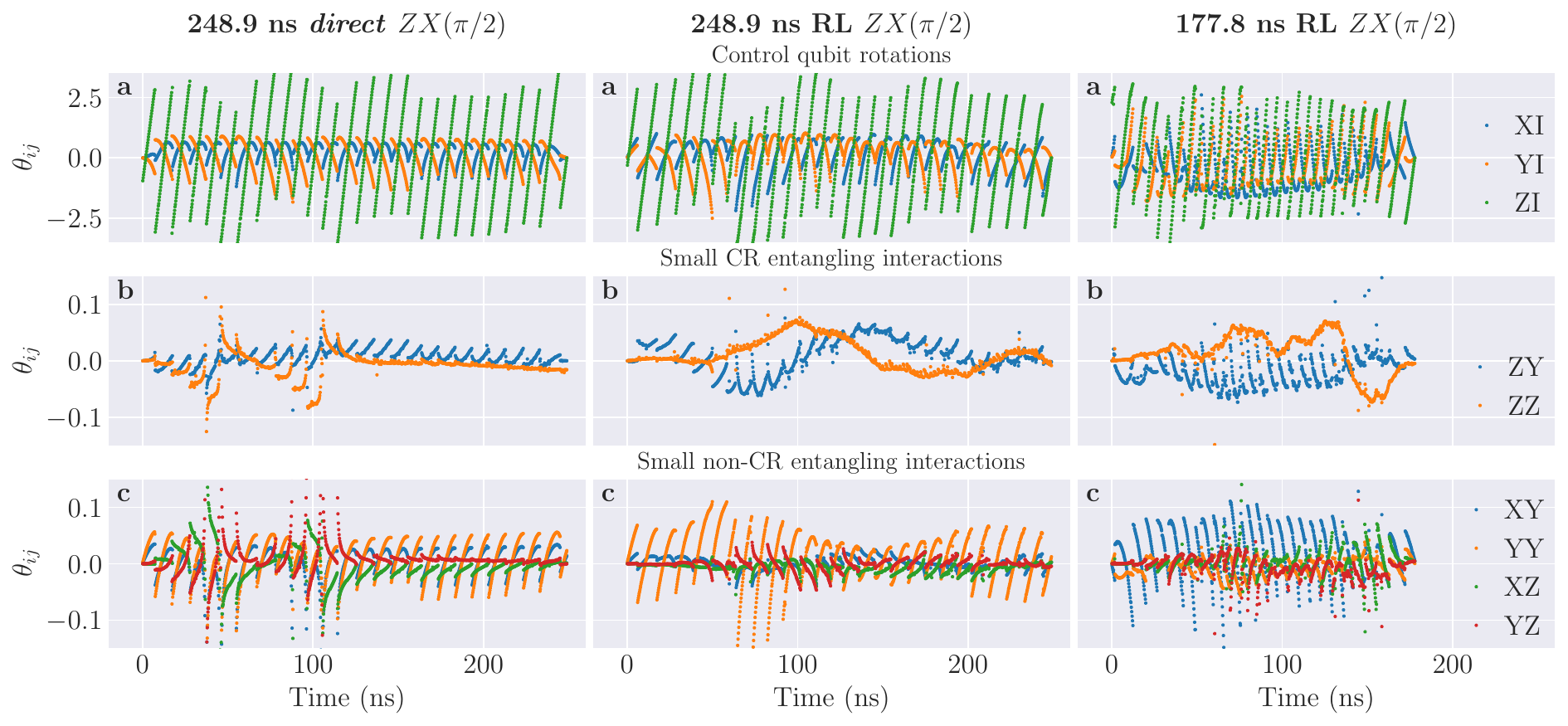}}
        \caption{\textbf{Remaining rotation angles of optimized $ZX(\pi/2$ pulses.} As complementary to Fig.~\ref{fig:entanglement_rotationangles} in the main text, we display the remaining rotation angles, categorized into: a) control qubit rotations, b) small entangling interactions expected from CR Hamiltonian, and c) small entangling interactions \emph{not} expected from CR Hamiltonian. Distinct evolution of rotation angles implies distinct physical processes in all three control solutions.}
        \label{fig:remaining_rotationangles}
\end{figure*}

Any two-qubit unitary map $U^{\rm qubit}(t,0)$ can be expressed in terms of a generating averaged Hamiltonian as follows
\begin{eqnarray}
    \exp(-iH^{\rm qubit}_{\rm avg}t) = \exp \left(-i\sum_{ij} \theta_{ij} \frac{P_i\otimes P_j}{2} \right),
\end{eqnarray}
where we have expanded $H^{\rm qubit}_{\rm avg}$ in the Pauli basis given by $P_i\in\{I,X,Y,Z\}$, and the rotation angle in the $ij$ direction depends on the duration $t$ and the $P_i\otimes P_j$ interaction strength. For example, the cross-resonance gate can be written as $ZX(\pi/2)=\exp(-i\pi ZX/4)$. 

For three-level transmons, we first compute the averaged Hamiltonian by taking the logarithm of the unitary $U(t,0)$, and then project it onto the qubit subspace. This allows us to quantify the strength of different interactions in the unitary at time $t$ via the rotation angle 
\begin{equation}
    \theta_{ij}(t) = \Tr\left[i\left(\Pi^{\rm qubit}\ln U(t,0)\Pi^{\rm qubit}\right)\frac{P_i\otimes P_j}{2}\right],
\end{equation}
where $\Pi^{\rm qubit}$ is the projector onto the qubit subspace. Computing the log of a matrix is non-trivial due to existence of branch cuts, which lead to different $\theta_{ij}$ from the same $U(t,0)$. To see this, we let $V$ be the unitary that diagonalizes $H_{\rm avg}$ and write
\begin{eqnarray}
    U(t,0) &=& \exp(-iH_{\rm avg}t) \nonumber \\
    &=& V\text{diag}(e^{-iE_1t},e^{-iE_2t},...)V^\dagger \nonumber \\
    &=& V\text{diag}(e^{-iE_1t-2in_1\pi},e^{-iE_2t-2in_2\pi},...)V^\dagger \nonumber \\
    &=& e^{-iV\text{diag}(E_1t+2n_1\pi,E_2t+2n_2\pi,...)V^\dagger } \nonumber \\
    \Rightarrow i \ln U(t,0)&=& V\text{diag}(E_1t+2n_1\pi,E_2t+2n_2\pi,...)V^\dagger, \nonumber \\
\end{eqnarray}
where we have taken into account the periodicity of the complex exponential via a list of integers $\{n_i\}$ corresponding to the eigenvalues $\{E_i\}$.  Note that a choice of $\{n_i\}$ specifies a particular branch cut where the \textit{principal} branch cut corresponds to $\{n_i = 0,  \forall i\}$.

Since we are most interested in the $ZX$ interaction, we would like to pick a branch cut where $\theta_{ZX}$ behaves nicely and without large jumps. With that goal in mind, we first consider the principal branch for a rough idea of how $\theta_{ZX}$ evolves. After that, we search through $\{n_i = \pm1\}$ at every time step, via brute force, to find the branch cuts that result in a smooth evolution of $\theta_{ZX}$, as seen in Fig.~\ref{fig:branchcuts}. These branch cuts can then be applied to obtain the evolution of rotation angle for the other interactions. Even though our approach is not perfect, which can be seen from the outliers in Fig.~\ref{fig:entanglement_rotationangles} and Fig.~\ref{fig:remaining_rotationangles}, the resultant data points are sufficiently accurate to reflect the main features of each evolution.

\section{Additional RL studies}

\subsection{RL optimization with 3 drives}\label{app:3drives}

\begin{figure}[t!]
    \centering
    \includegraphics[width=0.48\textwidth]{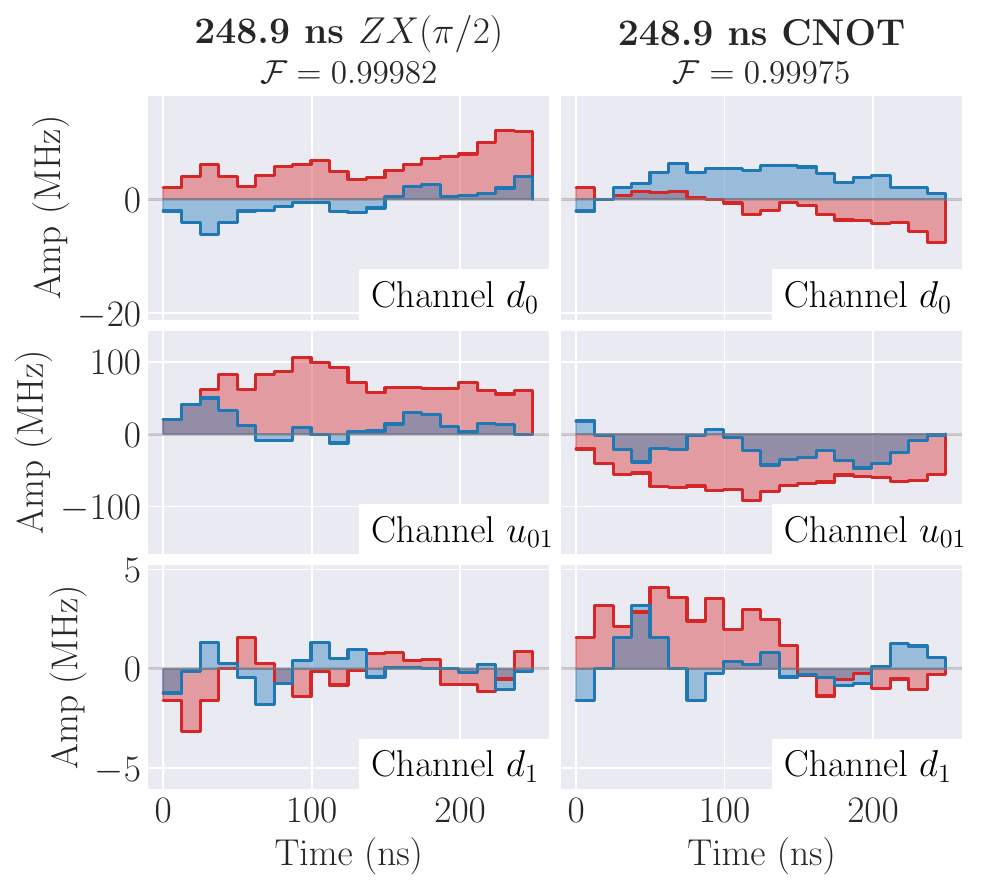}
    \caption{\textbf{RL optimized pulses for 3 drives.} With the single-qubit drive $d_0$ on the control transmon is included as compared to the main text, our DDPG RL agent effectively solves a 120-dimensional optimization problem. The control solutions found by our agent retain fidelity above 99.9\%. RL training hyperparameters are given in the ``3 drives" section of Table~\ref{tab:hyperparameters}. }
    \label{fig:3drives_pulse}
\end{figure}

\begin{figure}[t!]
    \centering
    \includegraphics[width=0.48\textwidth]{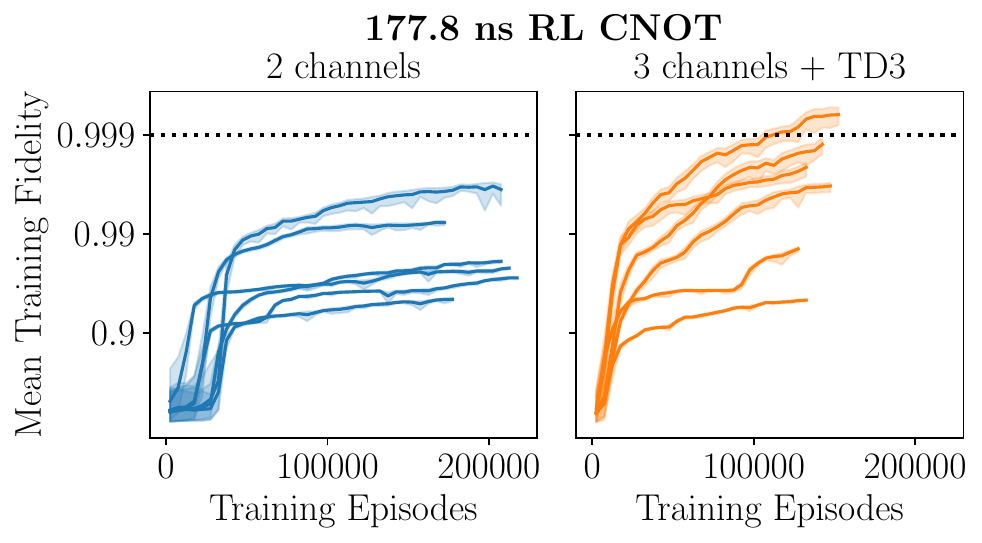}
    \caption{\textbf{Improvement when combining 3 drives and TD3.} Better learning curves as compared to employing DDPG with only 2 drives. RL training hyperparameters are given in the ``3 drives" section of Table~\ref{tab:hyperparameters}. }
    \label{fig:3drives_td3}
\end{figure}

Here we provide additional results when the agent has access to three drives $(d_0,u_{01},d_1)$. We first report the successful discovery of $\mathcal{F}\geq 99.9\%$ pulses by our RL agent using DDPG, effectively solving a 120-dimensional optimization problem. Examples of 248.9 ns pulses for implementing the $ZX(\pi/2)$ gates are shown in Fig.~\ref{fig:3drives_pulse}. Second, we observe an improvement in designing shorter gates when combining access to three drives with the TD3 tricks designed to stabilize training in this large action space. In Fig.~\ref{fig:3drives_td3}, we report learning curves for the task of designing a 177.8 ns CNOT pulse, obtained within the same run time. Despite slower training, TD3 on 3 drives exhibits an improved probability of successful runs. These additional findings suggest potential benefits when simultaneous control of all three drives is accessible.

\subsection{RL optimization with worst-case fidelity reward}\label{app:worstcase}

Here we summarize our investigation on using worst-case fidelity as an alternative figure of merit. Let us first discuss the standard approach of estimating worst-case fidelity over an ensemble of initial states restricted to qubit subspace. The restriction is valid as we focus on implementing quantum logic operations between two-level systems. Under this assumption, an arbitrary pure initial state can be written in terms of the computational basis as $\ket{\psi_0}=\sum_i c_i \ket i$ where $i\in\{0,1\}$ for one qubit and $i\in\{0,1,2,3\}$ for two qubits. The worst-case fidelity of a unitary map $U$ w.r.t $U_{\rm target}$ is defined as
\begin{eqnarray}
    \mathcal{F}_{\text{worst}} &=& \min_{\psi_0} \left|\ev{U^{\rm qubit}U^{\dagger}_{\rm target}}{\psi_0}\right|^2 \nonumber \\
    &=& \min_{\{c_i\}} \left|\sum_{ij}c_i^* c_j \mel{i}{U^{\rm qubit}U^{\dagger}_{\rm target}}{j}\right|^2, \label{eq:Fworst}
\end{eqnarray}
where $U^{\rm qubit}$ is the unitary map projected to the qubit subspace. The complex-valued coefficients $\{c_i\}$ can be recast into 3 (7) real values for one (two) qubit(s), where we have subtracted a global phase degree of freedom. Numerical optimization is then carried out via Sequential Least Squares Programming method (SLSQP), which we find to be the fastest and most stable out of all methods available in SciPy's library. It is should be emphasized that the worst-case fidelity can be estimated by simply evolving a few states initially in the computational basis, suggesting a straightforward implementation on near-term devices.

In fact, the estimation of the worst-case fidelity for a single qubit can be further improved by adopting a density matrix perspective. Working with a three-level system, a general qutrit density matrix is written as
\begin{eqnarray}
    \rho=\frac{1}{3}\left(I+\sqrt{3}\bm{r}\cdot\bm{\lambda}\right),
\end{eqnarray}
where $\bm{r}$  is an 8-dimensional Bloch vector for the qutrit state and $\bm{\lambda}$ is a vector of Gell-Mann matrices~\cite{generalbloch_ozols_mancinska}. The normalization condition for a pure state implies $|\bm{r}|=1$. Restricting the initial state to the qubit subspace leads to $r_i=0$ for $i=4,\dots,7$ and $r_8=1/2$, resulting in
\begin{eqnarray}
    \rho(0)=\frac{I}{3}+\frac{\lambda_8}{2\sqrt{3}} + \sum_{i=1}^3n_i\frac{\lambda_i}{2},
\end{eqnarray}
where we have rescaled the Bloch vector to the 3-dimensional unit sphere via $r_i=\sqrt{3}n_i/2$ so that $|\bm{n}|=1$. The relevant Gell-Mann matrices read
\begin{eqnarray}
    \lambda_i=
        \begin{pmatrix}
        \sigma_i &  \\
         & 0 
        \end{pmatrix}
    \text{ and }
    \lambda_8 = \frac{1}{\sqrt{3}}
        \begin{pmatrix}
        1& &  \\
         &1& \\
         & &-2
        \end{pmatrix},
\end{eqnarray}
where $\{\sigma_i\}$ denote the Pauli matrices. The fidelity of $\rho(0)$ evolved under a unitary $U=U(t,0)$ w.r.t.~a target unitary is
\begin{eqnarray}
    \mathcal{F} &=& \Tr \left[ U\rho(0)U^\dagger U_{\rm target}\rho(0)U^{\dagger}_{\rm target} \right] \nonumber \\
    &=& \frac{1}{3}+\frac{\Tr \left[\lambda_8(t)\lambda'_8 \right]}{12} + \sum_i n_i \frac{\Tr\left[\lambda_8(t)\lambda'_i+\lambda_i(t)\lambda'_8\right]}{4\sqrt{3}}  \nonumber \\
    &&+ \sum_{ij} n_i n_j \frac{\Tr\left[\lambda_i(t)\lambda'_j+\lambda_j(t)\lambda'_i\right]}{8} \nonumber \\
    &=& c +\sum_i n_i b_i + \frac{1}{2} \sum_{ij}n_i n_j A_{ij} \nonumber \\
    &=& \bm{n}^T\bm{b} + \frac{1}{2}\bm{n}^T(\bm{A}+2c\bm{I})\bm{n},
\end{eqnarray}
where we have symmetrized the quadratic term in the third line and used $\bm{n}^T \bm{n}=|\bm{n}|^2=1$ in the last. 

Evidently, minimizing fidelity over all possible initial qubit states is equivalent to \emph{minimizing a quadratic function over a sphere}. This spherically constrained quadratic programming problem (SCQP) can be efficiently solved using the algorithm outlined in Ref.~\cite{scqp_phanetal}. Indeed, the algorithm requires no initial guess, converges to a single solution within machine precision over multiple runs, and enjoys a $\sim10\times$ speed up compared to the standard SLSQP method. 

Finally, we note that a similar analysis for two qubits results in a quadratic programming problem for a 15-dimensional Bloch vector with highly non-trivial constraints beyond the normalization condition~\cite{qutritconstraint_gamel}, rendering the efficient SCQP algorithm inapplicable. Moreover,  optimizing for 15 parameters with multiple convoluted constraints turns out to be much harder and less stable than optimizing for 7 parameters as in Eq.~\ref{eq:Fworst}. Therefore, we deem the reparameterization unnecessary for two qubits and adhere to the standard approach using SLSQP algorithm. 

With the outlined methods, we train our RL agent to learn both single and two-qubit gates using worst-case fidelity as the figure of merit and find similarly high-fidelity control solutions. Due to the additional optimization, training with worst-case fidelity is slightly slower than training with average fidelity. Moreover, the uncertainty in its estimation using the SLSQP solver appears to destabilize training occasionally. Such an issue is fixed for learning a single-qubit gate when a more robust solver like SCQP is employed. Despite having no obvious advantage within this work, the worst-case fidelity remains an interesting alternative figure of merit to be further studied in future investigations.

% \nocite{*}

\bibliography{citation}% Produces the bibliography via BibTeX.

\end{document}